\documentclass[10pt,letterpaper]{article}
\usepackage[top=0.85in,left=2.75in,footskip=0.75in]{geometry}

\usepackage{amsmath,amssymb}

\usepackage{algorithm} 
\usepackage{algpseudocode} 

\usepackage{changepage}

\usepackage[utf8]{inputenc}
\usepackage[T1]{fontenc}

\usepackage{textcomp,marvosym}

\usepackage{cite}

\usepackage{nameref,hyperref}

\usepackage[right]{lineno}

\usepackage{microtype}
\DisableLigatures[f]{encoding = *, family = * }

\usepackage[table]{xcolor}

\usepackage{array}

\newcolumntype{+}{!{\vrule width 2pt}}

\newlength\savedwidth

\newcommand\thickhline{\noalign{\global\savedwidth\arrayrulewidth\global\arrayrulewidth 2pt}%
\hline
\noalign{\global\arrayrulewidth\savedwidth}}

\raggedright
\usepackage[aboveskip=1pt,labelfont=bf,labelsep=period,justification=raggedright,singlelinecheck=off]{caption}

\makeatletter
\renewcommand{\@biblabel}[1]{\quad#1.}
\makeatother

\usepackage{lastpage,fancyhdr,graphicx}
\usepackage{epstopdf}
\fancyheadoffset[L]{2.25in}
\fancyfootoffset[L]{2.25in}
\usepackage{subcaption}

\begin{document}
\vspace*{0.2in}

\begin{flushleft}
{\Large
\textbf\newline{Propagation of weakly advantageous mutations in cancer cell population} 
}
\bigskip
\newline

Andrzej Polanski\textsuperscript{1}, 
Mateusz Kania\textsuperscript{1}, 
Jarosław Gil\textsuperscript{1}, 
Wojciech Łabaj\textsuperscript{2}, 
Ewa Lach\textsuperscript{1}, 
Agnieszka Szczęsna\textsuperscript{1,*} 

\bigskip
\textbf{1} 
Department of Computer Graphics, Vision and Digital Systems, Silesian University of Technology, Akademicka 16, 44-100 Gliwice, Poland
\\
\textbf{2} Department of Data Science and Engineering, Silesian University of Technology, Akademicka 16, 44-100 Gliwice, Poland
\bigskip

%
%





* agnieszka.szczesna@polsl.pl

\end{flushleft}
\section*{Abstract}
Research into somatic mutations in cancer cell DNA and their role in tumour growth and progression between successive stages is crucial for improving our understanding of cancer evolution. Mathematical and computer modelling can provide valuable insights into the scenarios of cancer growth, the roles of somatic mutations, and the types and strengths of evolutionary forces they introduce. Previous studies have developed mathematical models of cancer evolution, incorporating driver and passenger somatic mutations. Driver mutations were assumed to have a strong advantageous effect on the growth of the cancer cell population, while passenger mutations were considered fully neutral or mildly deleterious. 

However, according to several studies, passenger mutations may have a weakly advantageous effect on tumour growth. In this paper, we develop models of cancer evolution with somatic mutations that introduce a weakly advantageous force to the evolution of cancer cells. The models used in this study can be classified into two categories: deterministic and stochastic. Deterministic models are based on systems of differential equations that balance the average number of cells and mutations during evolution. To verify the results of our deterministic modelling, we use a stochastic model based on the Gillespie algorithm. We compare the predictions of our modelling with some observational data on cancer evolution.


\section{Introduction}

Somatic mutations in cancer cells can be classified as either drivers or passengers, e.g., \cite{Graves_Maley_2012}. Driver mutations are rare but have a strong causal effect on cancer development, while passenger mutations are abundant in cancer cells DNA but have little or no effect on cancer development. Searching for driver mutations in cancer DNA and researching their roles in cancer growth plays a crucial role in understanding processes underlying oncogenesis, e.g., \cite{Campbell_et_al_2020}, \cite{Volgenstein_Kinzler_2015}. The number of driver mutations discovered/detected in a cancer cell population (in a sample from a cancer patient) is typically very low. These few drivers \cite{Volgenstein_Kinzler_2015} are always accompanied by large numbers of somatic passenger mutations. In contrast to driver mutations, impact of these numerous passenger mutations on cancer onset and/or development is presently rather not well understood \cite{Kumar_et_al_2020}. Due to its high significance the issue of the role of passenger mutations in cancer has been studied by using/combining experimental/clinical observations, repositories of cancer DNA sequencing data \cite{Campbell_et_al_2020}, \cite{CbioPortal}, \cite{COSMIC}, tools of comparative genomics \cite{Fu_Yao_2014}, statistics and mathematical modeling \cite{Bozic_et_al_2010}, \cite{Kumar_et_al_2020} \cite{McFarland_et_al_2013}. 

One viewpoint in the literature on cancer genomics is that all passenger mutations are fully neutral and have no effect on tumour progression/evolution. This opinion was presented in several studies \cite{Bozic_et_al_2010}, \cite{Williams_et_al_2016}, \cite{Tung_Durret_2021} supported by analyses of data from the ICGC/TCGA next-generation sequencing project \cite{Wilks_et_al_2014}. The rationale for the full neutrality hypothesis provided in \cite{Bozic_et_al_2010} was based on the consistency of predictions of the branching process model of cancer growth with allelic frequencies of driver and passenger mutations seen in sequencing data. Williams and colleagues \cite{Williams_et_al_2016} introduced a mathematical model for variant allele frequencies (VAF) in an exponentially growing population, which predicted the 1/f power law of distribution of VAFs of fully neutral passenger mutations, quite consistent with observational data of next-generation sequencing of cancer tissues. 
Some authors, \cite{McDonald_et_al_2018}, \cite{Noorbakhsh_et_al_2017}, \cite{Wang_et_al_2018}, pointed out that VAF statistics following from the model from \cite{Williams_et_al_2016} can also be reproduced by other models, with not necessarily fully neutral mutations. Tung and Durret \cite{Tung_Durret_2021} used the multi-type branching processes model for studying the possibility of distinguishing neutral from advantageous mutations. 

Many researches, however, have been bringing evidence that accumulated passenger mutations can impact cancer evolution, parallelly/additionally to drivers. Several authors observed that statistics (patterns of allelic frequencies) of passenger mutations differ between different cancers. E.g., Jjao and coauthors \cite{Jiao_et_al_2020} and Salvadores and coauthors \cite{Salvadores_et_al_2019} demonstrated that genomic locations and frequencies of somatic mutations can be used to construct molecular signatures to distinguish between cancer types and their progression scenarios. In the recent research \cite{Kumar_et_al_2020} the authors used quantitative molecular functional impact score \cite{Fu_Yao_2014} and evolutionary conservation measure \cite{Davydov_et_al_2010} to quantify the cumulative fitness effects of passenger mutations on tumour growth. Applying these scores to data on pan-cancer whole genome sequencing \cite{Campbell_et_al_2020} they concluded that aggregated effect of passengers plays a role in tumorigenesis beyond standard drivers and may either introduce weakly deleterious impact or may generate mildly driving (advantageous) evolutionary force. McFarland and coauthors \cite{McFarland_et_al_2013}, \cite{McFarland_et_al_2014} analyzed somatic mutations available in Cosmic database \cite{COSMIC} and, by studying their potential molecular impact using the bioinformatic tool PolyPhen \cite{Boyko_et_al_2008}, hypothesized that majority of the passenger mutations are likely to exert a mildly damaging effect on the evolution of cancer cells population. Following this hypothesis, they have developed a mathematical model of cancer evolution with two counteracting factors, frequent passenger mutations each with a weak deleterious effect and rare, driver mutations. Accumulated passenger mutations caused a slow shrinking of the cancer population, while rare driver mutations introduced selective sweeps, i.e., short time intervals of rapid population growth. 

As outlined above, mathematical/computational modelling of tumour evolution involved scenarios of evolution with driver mutations accompanied by fully neutral passengers \cite{Bozic_et_al_2010}, \cite{Williams_et_al_2016} and driver mutations interacting with mildly deleterious passenger mutations \cite{McFarland_et_al_2013}, \cite{McFarland_et_al_2014}. 
However, despite substantial experimental evidence on occurrence of weakly advantageous mutations in cancer cells, models of tumor progression driven by weakly advantageous somatic mutations were not researched or confronted against available observational data. In this paper, we elaborate and analyse mathematical and computational models of oncogenesis driven by weakly advantageous passenger mutations and we compare their predictions with some experimental data. 
Scenario and contributions of our analysis are as follows:
\begin{itemize}
    \item We formulate deterministic model of tumor evolution as a system of differential balance equations for changes of expected numbers of cells with divisions and deaths, and for changes of expected numbers of occurring mutations. 
    \item In this model of tumor growth solitary mutation wave propagates in cancer cells population. We obtain analytical relations concerning the propagation of the mutation wave (dynamics of mutation wave) in the growing population of cancer cells, in terms of mean and variance of numbers of mutations. In the previous literature, models of dynamics of mutation of fitness waves were derived under the assumption of constant population size.
    \item We elaborate and launch stochastic simulations for scenarios of cancer evolution with weakly advantageous mutations based on the Gillespie algorithm, for supporting and verifying results of deterministic modelling.
    \item We analyse the model of quasi - stationary mutation wave (mutation wave with variance approximately constant over time). Using the deterministic model augmented with the simple cutoff condition for number of cells \cite{Tsimring_et_al_1996}, \cite{Kessler_et_al_1997} we establish deterministic numerical procedure for relating the variance of the quasi - stationary mutation wave with the population size. For the model of quasi - stationary mutation wave we derive analytical relations for growth rate of cancer cells population. We verify obtained deterministic, analytical results by stochastic simulations.
    \item We confront/support both obtained analytical results and simulation results with some of the publicly available experimental data. We show data / studies that can support the hypothesis of tumor growth driven solely by weakly advantageous somatic mutations. 
\end{itemize}

\section{Methods}

Methods applied in this paper include:
\begin{itemize}
\item deterministic modelling of propagation of mutations in cancer cell populations, 
\item stochastic model based on Gillespie engine for comparison with results of deterministic modelling,
\item bioinformatic data analysis pipeline for calling somatic mutations in publicly available single-cell DNA sequencing experimental data of breast cancer cells. 
\end{itemize}

\begin{figure}[ht]
\centering
\includegraphics[width=0.8\textwidth]{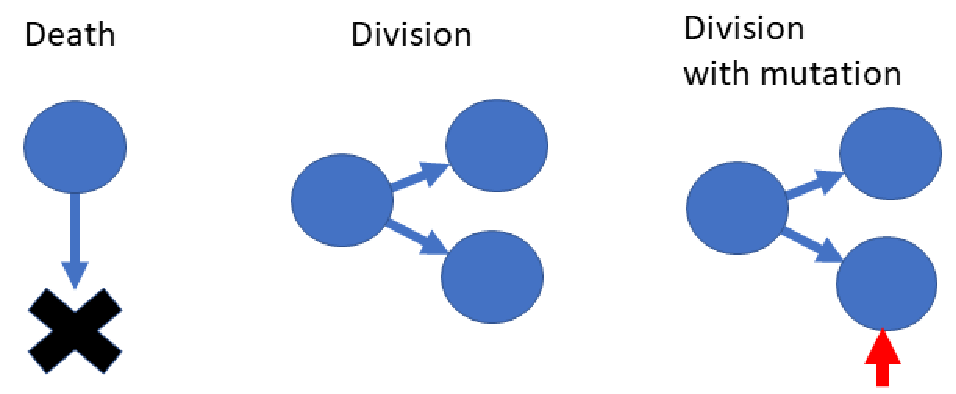}
\caption{{\bf Graphical representation of possible events in the analyzed scenario of cancer cells populations evolution.}
}
\label{fig1}
\end{figure}

We study evolution scenario of the cancer cells population with events of deaths, births and weakly advantageous mutations. These events are represented graphically in  Fig~\ref{fig1}. Cell deaths and births are inhomogeneous Poisson processes with rates depending on the state of the model (defined by population size and the number of mutations in cancer cells). Random events of mutations are occurring during cell births. Cell death rate depends on the cancer cell population size, on the population capacity parameter and on the exponent parameter. Birth rates of cells depend on the number of (somatic, weakly advantageous) mutations in their DNA.  

\subsection{Notations for processes and events}

In relation to the events shown in  Fig~\ref{fig1} we use the following notation. 

$N(t)$ denotes population size, i.e., the expected number of cells in the analyzed cancer population and by $N_{C}$, we denote the population capacity parameter. We often drop $t$ (time) writing $N(t)=N$, for conciseness. The intensity of death processes is assumed to depend on the population size $N(t)$ and population capacity parameter $N_{C}$ by a power function with exponent denoted by $A$ (see  Eq~(\ref{eq:death_intensity})).

By $l$ we denote the number of weakly advantageous passenger mutations in a cancer cell, $n_{l}(t)=n_{l}$ stands for the size (expected number of cells) of the subpopulation of cancer cells population, of cells harbouring $l$ weakly advantageous passenger mutations. We also call the sub-population $n_{l}$ - type or class with $l$ mutations. By $f$ we denote the value of the positive selection coefficient. The probability of acquiring a mutation in a cell division is denoted by $p_{f}$.

All variables and parameters presented above are listed in Table~\ref{table_1} below.

\begin{table}[!ht]
\centering
\caption{
{\bf Variables and parameters in deterministic and stochastic
models and their explanations.}}
\begin{tabular}{|c|l|}
\hline
{\bf Symbol} & {\bf Explanation}         \\
\thickhline
$N(t)$            & cancer cells population size      \\ \hline
$N_{C}$             & population capacity parameter      \\ \hline
$l$                & number of weakly advantageous mutation in cancer cell \\ \hline
$n_{l}(t)$       & size of the subpopulation of cells with l mutations  \\ \hline
$f$                & value of positive selection coefficient    \\ \hline
$p_{f}$             & probability of acquiring mutation in cell division  \\ \hline
$A$                & exponent of the power function describing cell death intensity
\\ \hline
\end{tabular}
\label{table_1}
\end{table}

\subsubsection{Death process}

Intensity of the cellular death Poisson processes $\mu_{D}(N)$ depends only on the total number of cells $N$ and on the population capacity parameter $N_{C},$ by the relation
\begin{equation}
\mu_{D}(N)=\left(\frac{N}{N_{C}}\right)^{A}.\label{eq:death_intensity}
\end{equation}
The above relation between $\mu_{D}(N)$ and $N$ is assumed as a power function with exponent $A$. If the exponent is equal to one, $A=1,$ the model of the death intensity becomes the same as that used in the literature, \cite{McFarland_et_al_2013}, \cite{McFarland_et_al_2014}. Here, by allowing $A\neq1$ we introduce one more degree of freedom in modelling. This additional degree of freedom has consequences to the rate of growth of cancer cells population with weakly advantageous mutations and allows us to fit our model to recently published experimental results on rates of growth of tumours \cite{PerezGarcia_et_al_2020}.

\subsubsection{Birth process}

Birth rate of cells of type $l$, denoted by $\mu_{B}(l,f)$, is assumed to be described by the following function

\begin{equation}
\mu_{B}(l,f)=(1+f)^{l}\simeq e^{lf}\label{eq:birth_fit}
\end{equation}
Accumulation of weakly advantageous mutations, according to the above function, results in an increase in the intensity of the birth process. On the basis of the assumption of the weak effect of mutations, we can linearize the birth rate function Eq~(\ref{eq:birth_fit}) for approximating values of $\mu_{B}(l,f)$

\begin{equation}
\mu_{B}(l,f)\cong\mu_{B}(\chi_{f},f)(1+f(l-\chi_{f})),\;\mu_{B}(l-1,l)\cong\mu_{B}(\chi_{f},f)(1+f(l-1-\chi_{f})).\label{eq:linearized_fit}
\end{equation}
The above linearization is around the mean value of the number of mutations in a cell, $\chi_{f}$, defined in the subsequent text in Eq~(\ref{eq:mean_l_def_fit}).

\subsubsection{Mutation process}

Mutations occur during cell divisions, with probability $p_{f}$, which leads to the intensity of mutations given by

\begin{equation}
p_{f}\mu_{B}\label{eq:muts_inten_fit}
\end{equation}
where $\mu_{B}$is birth process intensity given by Eq~(\ref{eq:birth_fit}).

\subsubsection{Units of time scale}
We define the units of time scale for reporting results of computations and simulations by using the intensity of birth process of the "wild type" cells (cells with no mutations).
In case of no mutations present in cells, from Eq~(\ref{eq:birth_fit}) we have $\mu_{B}(0,f)=1$. This is used to scale / measure the time lapse. The unit of time is assumed equal to the expected waiting time for the birth event in a "wild type" cell, with no mutations ($l=0$). 

\subsubsection{Values of parameters}

Values of parameters of selection and mutation processes in the model are selected by referring to the literature \cite{McFarland_et_al_2013}. The value of the negative selection coefficient of weakly deleterious passenger mutations was taken by these authors to be in the range $10^{-1}-10^{-4}$. We take these values as a reference, and we assume a positive selection coefficient, $f$, of the order $10^{-4}-10^{-3}$. Probability of mutations, $p_{f}$ are taken to be of the order of $10^{-3}-10^{-2}$(in \cite{McFarland_et_al_2013} similar values of probability of passenger mutations are obtained by multiplying mutation intensity per cell division event per nucleotide by the estimated number of target sites equal to be of the order of $10^{6}$). The population size of cancer cells in our computations is assumed in the range from $10^{3}$ to $10^{6}$. 

\subsection{Deterministic modelling of propagation of mutations and evolution
of the population of cancer cells}

Deterministic modelling involves formulating systems of differential equations describing the fitness effects of advantageous mutations as well as related laws behind cellular deaths and replications. 
On the basis of relations Eq~(\ref{eq:death_intensity})-Eq~(\ref{eq:birth_fit}) we formulate a deterministic model of the evolution as the set of deterministic equations of balances of expected streams of dividing/dying cells and occurring passenger mutations. The set (system) of equations of balances of cells/mutations flows has the following form

\begin{equation}
\frac{d}{dt}n_{l}=p_{f}\mu_{B}(l-1,f)n_{l-1}+(1-p_{f})\mu_{B}(l,f)n_{l}-\mu_{D}(N)n_{l}.\label{eq:master_fit}
\end{equation}
The right-hand side has three components of the rate of change of a number of cells of type $l$. The first component is the rate of increase of the number of cells of type $l$ due to cells of type $l-1$ acquiring mutation during their replications, the second component gives the rate of increase due to replications of cells of type $l$ (division without mutation), finally, the third one is the rate of decrease due to cell deaths. The range of indices $l$ is $l=0,1,2\ldots$

\subsubsection{Evolution of the population size \label{subsec:Evolution-of-the_population_size}}

We can derive a differential equation governing the evolution of the population size - the total number of cells $N$ equal to the sum of the numbers of cells over all cell types. We have $N=\sum_{l}n_{l}$ and we define mean $\chi_{f}$ and variance $\sigma_{f}^{2}$ of numbers of advantageous mutations

\begin{equation}
\chi_{f}(t)=\chi_{f}=\sum_{l}l\nu_{l},\label{eq:mean_l_def_fit}
\end{equation}

\begin{equation}
\sigma_{f}^{2}=\sum_{l}(l-\chi_{f})^{2}\nu_{l},\label{eq:var_l_def_fit}
\end{equation}
where $\nu_{l}$ are frequencies of cells of type $l$,

\begin{equation}
\nu_{l}(t)=\frac{n_{l}(t)}{N(t)}.\label{eq:freq_fit}
\end{equation}
Summing both sides of equations Eq~(\ref{eq:master_fit}) over the range of values of $l$ and using linear approximations Eq~(\ref{eq:linearized_fit}) and relation Eq~(\ref{eq:death_intensity}) we get 

\begin{equation}
\frac{d}{dt}N=\left[\mu_{B}(\chi_{f},f)-\left(\frac{N}{N_{C}}\right)^{A}\right]N,\label{eq:pop_size_eq_fit}
\end{equation}
The above differential equation describes the dynamics of the size of the total cell population $N(t)$. 
Rewriting Eq~(\ref{eq:pop_size_eq_fit}) as

\begin{equation}
\frac{1}{N} \frac{d}{dt}N=\left[\mu_{B}(\chi_{f},f)-\left(\frac{N}{N_{C}}\right)^{A}\right],\label{eq:pop_size_eq_fit_fast}
\end{equation}
and noting that $N$ is large one can notice that Eq~\ref{eq:master_fit} with Eq~\ref{eq:pop_size_eq_fit_fast}
can be considered as a two time scale system \cite{Kuehn_2015}.  
In the fast time scale we assume $\mu_{B}(\chi_{f},f)\simeq const$ so starting from any initial condition the population size $N(t)$ tends to the fast time limit $N=N_{C}\mu_{B}(\chi_{f},f)^{\frac{1}{A}}$. Slow time scale dynamics is given by solution to 
Eq~\ref{eq:master_fit} restricted to the manifold
\begin{equation}
N(t)=N_{C}\left[\mu_{B}(\chi_{f}(t),f)\right]^{\frac{1}{A}}.\label{eq:N_slow_fit}
\end{equation}

\subsubsection{Mutation wave \label{subsec:Evolution-of-mean}}

We describe mutation wave traveling in cancer cells population by deriving dynamics of the change of mean number of mutations $\chi_{f}(t)$. In order to derive equations describing the change of mean numbers of mutations over time we start by considering frequencies of cell types. We first differentiate both sides of Eq~(\ref{eq:freq_fit}) to obtain

\begin{equation}
\frac{d}{dt}\nu_{l}=-\frac{1}{N^{2}}\frac{dN}{dt}n_{l}+\frac{1}{N}\frac{dn_{l}}{dt}.\label{eq:freq_deriv_fit}
\end{equation}
By substituting Eq~(\ref{eq:pop_size_eq_fit}) and Eq~(\ref{eq:master_fit}) in the above equation we get the set of differential equations describing dynamics of $\nu_{l}(t)$

\begin{equation}
\frac{d}{dt}\nu_{l}=\mu_{B}(\chi_{f},f)\left[p_{f}(1+f(l-1-\chi_{f}))\nu_{l-1}-p_{f}(1+f(l-\chi_{l}))\nu_{l}+f(l-\chi_{f})\nu_{l}\right].\label{eq:freqs_dyn_fit}
\end{equation}

Using the above, we derive a differential equation describing the time change of $\chi_{f}(t)$. We use equations for dynamics of frequencies of cell types, Eq~(\ref{eq:freqs_dyn_fit}). Multiplying both sides of Eq~(\ref{eq:freqs_dyn_fit}) by $l$ and summing up over the range of index $l$ we obtain

\begin{equation}
\sum_{l}l(\frac{d}{dt}\nu_{l})=\frac{d}{dt}(\chi_{f})=\mu_{B}(\chi_{f},f)(p_{f}+f\sigma_{f}^{2}).\label{eq:wave_fit}
\end{equation}

\subsubsection{Quasi - stationary profile of mutation waves in finite - size population\label{subsec:Finite-poplation-size}}

Eq~(\ref{eq:wave_fit}) is used here for modeling propagation of mutation wave. However, ist efficient application requires elaborating a method for computing/estimating values of variance $\sigma_{f}^{2}$. Balancing forces, which influence values of $\sigma_{f}^{2}$ needs accounting for the effects of the finite size of the cancer cell population and and quantization of sub-populations (cell classes) $n_l$. Differential balance equations  Eq~(\ref{eq:master_fit}) are formulated in real numbers arithmetic, so they contain streams of cells and mutations generated in sub-populations $n_l$ of fractional sizes (of sizes smaller than $1$). In real cellular populations and in stochastic simulations sub-populations $n_l$ generate cells and mutations when their size is bigger than $1$.
So here we use a simple (heuristic) modification of the deterministic model Eq~(\ref{eq:master_fit}), which involves introducing the assumption that cell divisions in the cell class $n_{l}$ can happen only if $n_{l}\geq1$. 
Introducing the function $h(n)$ defined as
\begin{equation}
h(n)=\left\{ \begin{array}{c}
n\;\mathrm{if}\;n\geqslant1\\
0\;\mathrm{if}\;n<1
\end{array}\right..\label{eq:FP_modifications}
\end{equation}
we formulate modified deterministic balance equations, as follows

\begin{equation}
\frac{d}{dt}n_{l}=p_{f}\mu_{B}(l-1,f)h(n_{l-1})+(1-p_{f})\mu_{B}(l,f) h(n_{l})-\mu_{D}(N)n_{l}.\label{eq:master_fit_FPmod}
\end{equation}

The condition for the size of cell classes analogous to Eq~(\ref{eq:FP_modifications}) was already used in modeling evolution of fitness of RNA viruses in populations with constant size where it was named cutoff condition \cite{Tsimring_et_al_1996}, \cite{Kessler_et_al_1997}. 
It is intuitively explained by the same argument as we are giving here and
additionally supported by experimental and simulation results.
Here we use the cutoff modification for the model with growing population size. Analogously to the previous literature we observe that deterministic model with cutoff condition Eq~(\ref{eq:master_fit_FPmod}) gives reasonably good consistency with stochastic simulations.

\begin{figure}[ht]
\includegraphics[width=1\textwidth]{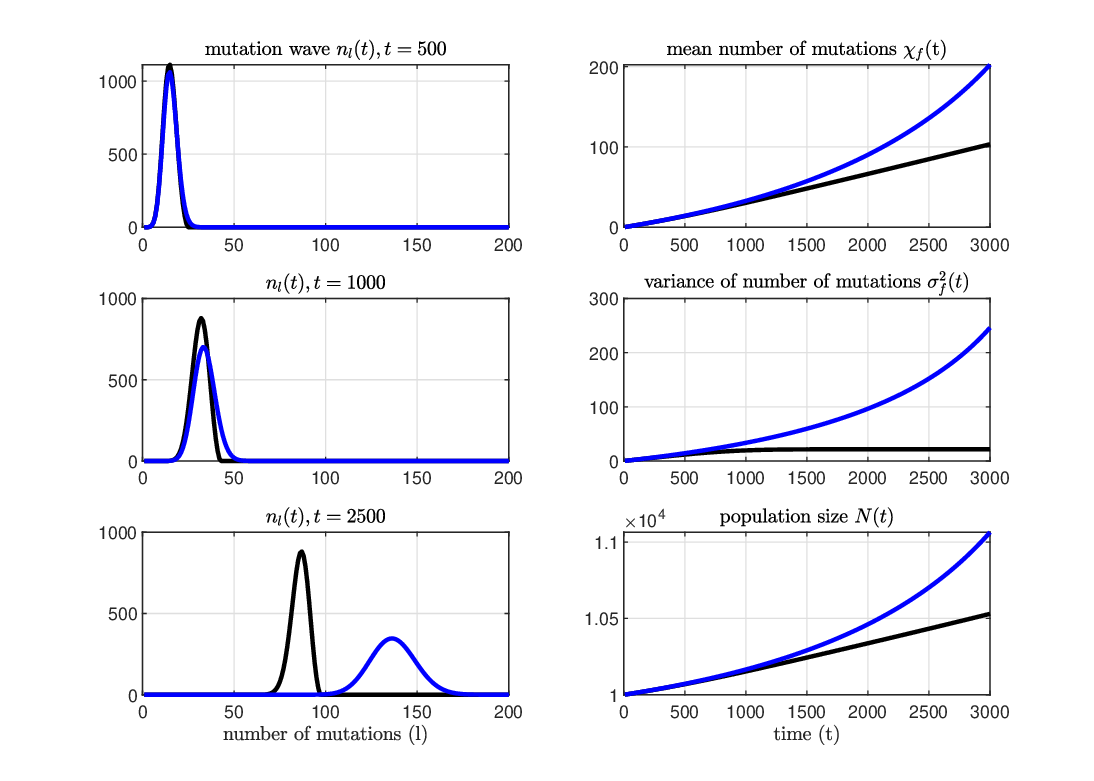}
\caption{{\bf }
Comparison of solutions to deterministic differential equations model Eq~(\ref{eq:master_fit}), with the solution to model Eq~(\ref{eq:master_fit_FPmod}) with cutoff modification Eq~(\ref{eq:FP_modifications}), for modelling evolution with weakly advantageous mutations, for the parameter, set $A=1$, $f=0.0005$, $p_{f}=0.025$, $N_{C}=10~000$. Left panels present plots of mutation waves at three time instants $t=500$, $t=1~000$ and $t=2~500$. The right panels show time plots of mean numbers of mutations, $\chi_{f}(t),$ (upper plots), the variance of mutation numbers, $\sigma_{f}^{2}(t)$, (middle plots) and number of cells in the cancer population $N(t)$ (lower plots). For all plots, bold blue, solid lines show time plots computed by using numerical integration of the system of differential equations without Eq~(\ref{eq:master_fit}) while analogous black lines represent solutions to modified equations, Eq~(\ref{eq:master_fit_FPmod}).  
}
\label{fig2}
\end{figure}

 Comparison of numerical solutions to two models, differential system of differential equations Eq~(\ref{eq:master_fit}) and modified differential equations with cutoff condition Eq~(\ref{eq:master_fit_FPmod}) is given in  Fig~\ref{fig2}, for the parameter set $A=1$, $f=0.0005$, $p_{f}=0.025$, $N_{C}=10~000$. Time plots corresponding to solutions
to (\ref{eq:master_fit}) are drawn in blue, while those representing Eq~(\ref{eq:master_fit_FPmod}), are drawn in black. Initial conditions for both models are defined by initial population size $N(0)=N_{C}$ and the initial number of mutations in all cells $l=0$. Numerical solutions are computed by the fourth-order Runge-Kutta algorithm. Qualitatively, both solutions are analogous, both predict the propagation of mutation wave towards accumulating increasing number of weakly advantageous passenger mutations by cancer cells and the increase of the population size $N(t)$. However, quantitatively the two scenarios of propagation differ significantly. Mutation wave computed on the basis of equations, Eq~(\ref{eq:master_fit}) propagates faster. When represented by means and variances it is also broader (has larger variance) than that corresponding to  Eq~(\ref{eq:master_fit_FPmod}). When comparing time plots of variances of mutation numbers, $\sigma_{f}^{2}(t)$ (right panel, middle plot), we see that the values of $\sigma_{f}^{2}(t)$ computed based on cutoff modified equations Eq~(\ref{eq:master_fit_FPmod}) (black line) are approximately constant (precisely, they increase very slowly), while values of $\sigma_{f}^{2}(t)$ computed on the basis differential equations Eq~(\ref{eq:master_fit}) (blue line) show a significant increase in time.

\begin{figure}[ht]
\includegraphics[width=1\textwidth]{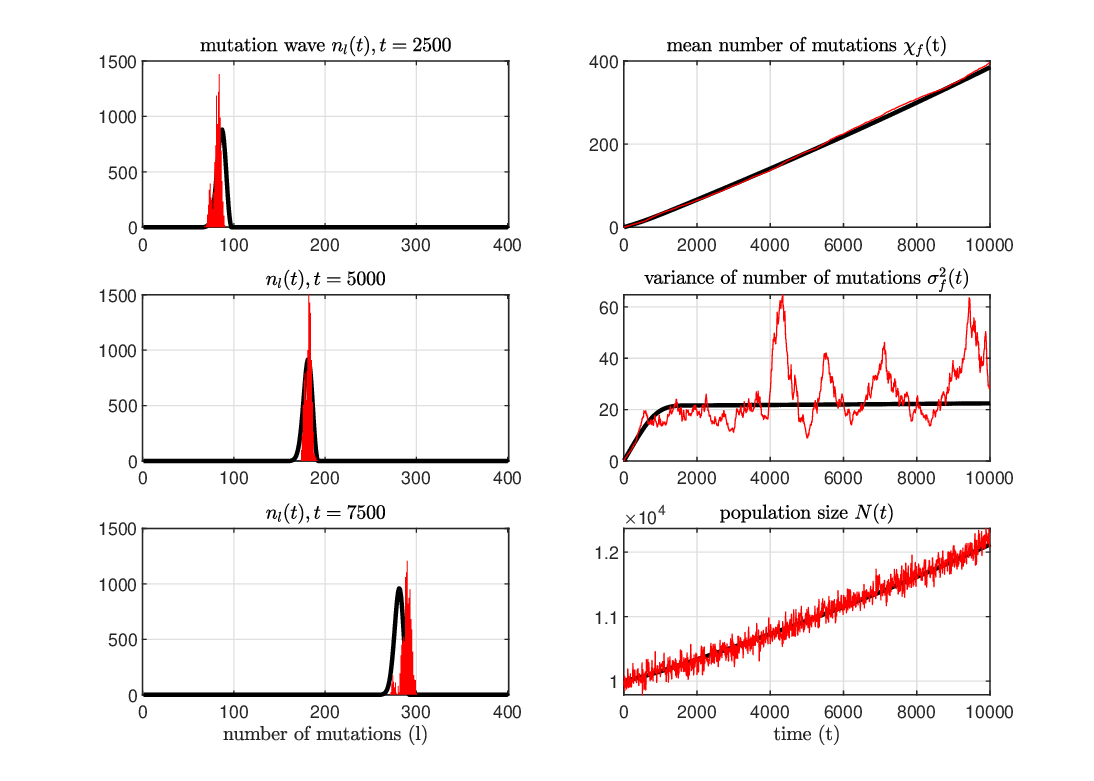}
\caption{{\bf }
Comparison of deterministic (with cutoff modification) versus stochastic modelling for
evolution with weakly advantageous mutations, for the parameter set $A=1$,
$f=0.0005$, $p_{f}=0.025$, $N_{C}=10~000$. Left panels present plots
of mutation waves at three time instants $t=2~500$, $t=5~000$ and
$t=7~500$. The right panels show time plots of mean numbers of mutations,
$\chi_{f}(t),$ (upper plots), the variance of mutation numbers, $\sigma_{f}^{2}(t)$,
(middle plots) and the number of cells in the cancer population $N(t)$
(lower plots). For all plots, black, bold, solid lines show time plots
computed by using numerical integration of the system of differential
equations with cutoff modification Eq~(\ref{eq:master_fit_FPmod}).
Red plots present the results of stochastic simulations obtained by using
Gillespie algorithm.
}
\label{fig3}
\end{figure}

Additionally it can be seen that solutions to modified differential equations Eq~(\ref{eq:master_fit_FPmod}) are consistent with the results of stochastic simulations. This can be demonstrated by comparison analogous to that presented in  Fig~\ref{fig2}. Deterministic modelling is based on differential equations with cutoff modification. Stochastic simulations are based on the Gillespie algorithm described in the forthcoming subsection.
The set of parameters assumed in computations is the same as that in  Fig~\ref{fig2}, $A=1$, $f=0.0005$, $p_{f}=0.025$, $N_{C}=10~000$. Here we assume a longer time range, from $t=0$ to $t=10~000$. In  Fig~\ref{fig3} we show time plots corresponding to differential equations with cutoff modification Eq~(\ref{eq:master_fit_FPmod}) solved numerically by using Runge-Kutta method, versus analogous time plots obtained by using stochastic simulations with Gillespie algorithm. The deterministic differential equation model and Gillespie simulation algorithm were started with the initial condition given by initial population size $N(0)=N_{C}$ and the initial number of mutations in all cells $l=0$.

In the plots of time change of $\sigma_{f}^{2}(t)$ corresponding to solutions of
systems of cutoff modified equations Eq~(\ref{eq:master_fit_FPmod}) shown in in Fig~\ref{fig2}
and Fig~\ref{fig3} we observe quasi - stationarity of $\sigma_{f}^{2}(N)$.
Variances of mutation numbers $\sigma_{f}^{2}(t)$ corresponding to propagating mutation waves are changing slowly in time. It is also the property of mutation waves seen stochastic in simulations, in  Fig~\ref{fig3} (when variances of numbers of mutations are averaged over time).

\begin{figure}[ht]
\includegraphics[width=1\textwidth]{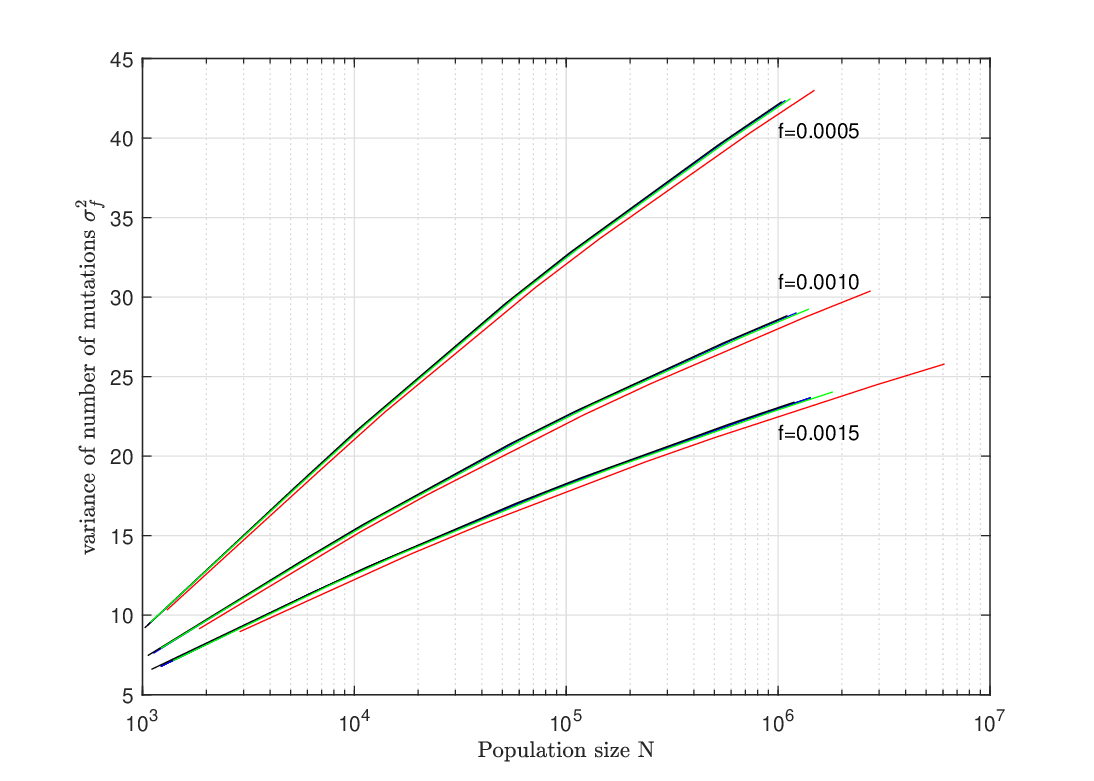}
\caption{{\bf }
plots of functions $\sigma_{f}^{2}(N)$ for different values of positive selection coefficient ($f=0.0005$, $f=0.001$, $f=0.0015$) and for different values of exponent parameter ($A=0.1$, $A=0.3$, $A=0.5$, $A=1$). Plots corresponding to different values of $A$ are drawn with different colours. $A=0.1$ - with red color, $A=0.3$ - with green color, $A=0.5$ - with blue color and $A=1.0$ - with black color.
}
\label{fig4}
\end{figure}

Referring to the slow time relation Eq~\ref{eq:N_slow_fit} we accept the hypothesis that $\sigma_{f}^{2}$ is a slowly changing function of the population size $N$. Here we present the slow change of $\sigma_{f}^{2}(N)$ in quantitative terms. In  Fig~\ref{fig4} we show plots of functions $\sigma_{f}^{2}(N)$ for different values of positive selection coefficient ($f=0.0005$ ,$f=0.001$ , $f=0.0015$) and for different values of exponent parameter ($A=0.1$, $A=0.3$, $A=0.5$, $A=1$). Plots in  Fig~\ref{fig4} were obtained by multiple runs of a numerical algorithm for solving cutoff modified equations, Eq~(\ref{eq:master_fit_FPmod}), with different values of $N_{C}$. It is seen from  Fig~\ref{fig4} that, while values of $N$ change in the range of $6$ orders of magnitude (from $10^{3}$to $10^{6}$) the corresponding values of $\sigma_{f}^{2}$ span less than one order of magnitude (from $10$ to $50$). Plots corresponding to different values of $A$ are drawn with different colours.  Fig~\ref{fig4}, apart from showing slow growth $\sigma_{f}^{2}(N)$ as a function of $N$ also demonstrates that changing the value of the exponent $A$ makes almost no change in the plots of $\sigma_{f}^{2}(N)$.

Based on the above, in the model considered in the next subsection (for deriving the power law of the population size growth) we use approximation 

\begin{equation} \sigma_{f}^{2}(N)=\sigma_{f}^{2}\cong const.\label{eq:sigma_square_fit_const}
\end{equation}

\subsubsection{Power law in the evolution of the population size \label{subsec:Evolution-in-the-slow_time}}

Evolution of the size of cancer cells population in the slow time with weakly advantageous mutations Eq~(\ref{eq:N_slow_fit}) is determined by the time change of mean numbers of mutations in cancer cells, $\chi_{f}(t)$, whose dynamics is described by differential equations Eq~(\ref{eq:wave_fit}). The equation for the velocity of mutation wave Eq~(\ref{eq:wave_fit}) can be used to derive a model of the evolution of the population size in the slow time in the form of differential equations.
Through this subsection, we accept the hypothesis on the slow change of variance, Eq~(\ref{eq:sigma_square_fit_const}).

To derive a differential equation for the slow time evolution of the population size, we differentiate with respect to time both sides of Eq~(\ref{eq:N_slow_fit}), and we use Eq~(\ref{eq:wave_fit}). This leads to

\[
\frac{d}{dt}N(t)=\frac{1}{A}N_{C}\left[\mu_{B}(\chi_{f},f)\right]^{(1+\frac{1}{A})}(fp_{f}+f^{2}\sigma_{f}^{2}).
\]

The above equation, by using Eq~(\ref{eq:N_slow_fit}) can be transformed to the differential equation with $N(t)$ as state
variable,
\begin{equation}
\frac{d}{dt}N(t)=\frac{1}{AN_{C}^{A}}N^{(1+A)}(fp_{f}+f^{2}\sigma_{f}^{2}).\label{eq:dtN_fit_N}
\end{equation}

By computing logarithms of both sides of the above equation, we have the relation

\begin{equation}
log_{10}(\frac{d}{dt}N(t))=(1+A)log_{10}(N(t))+log_{10}\left(\frac{fp_{f}+f^{2}\sigma_{f}^{2}}{AN_{C}^{A}}\right),\label{eq:log_dtN_fit_N}
\end{equation}
which represents the power law in the population size growth. Logarithms of population size $log_{10}(N(t))$ and the growth rate $log_{10}(\frac{d}{dt}N(t))$ are dependent linearly with the coefficient $1+A$. Analytical solution to the differential equation Eq~(\ref{eq:dtN_fit_N}),

\begin{equation}
N(t)=\frac{N_{C}}{\left[1-t(fp_{f}+f^{2}\sigma_{f}^{2})\right]^{\frac{1}{A}}}\label{eq:N_analytical}
\end{equation}
exhibits a finite time escape at $t=\frac{1}{fp_{f}+f^{2}\sigma_{f}^{2}}$. 

\begin{figure}[ht]
\centering
 \begin{subfigure}{0.7\textwidth}
 \caption{}
\includegraphics[width=\linewidth]{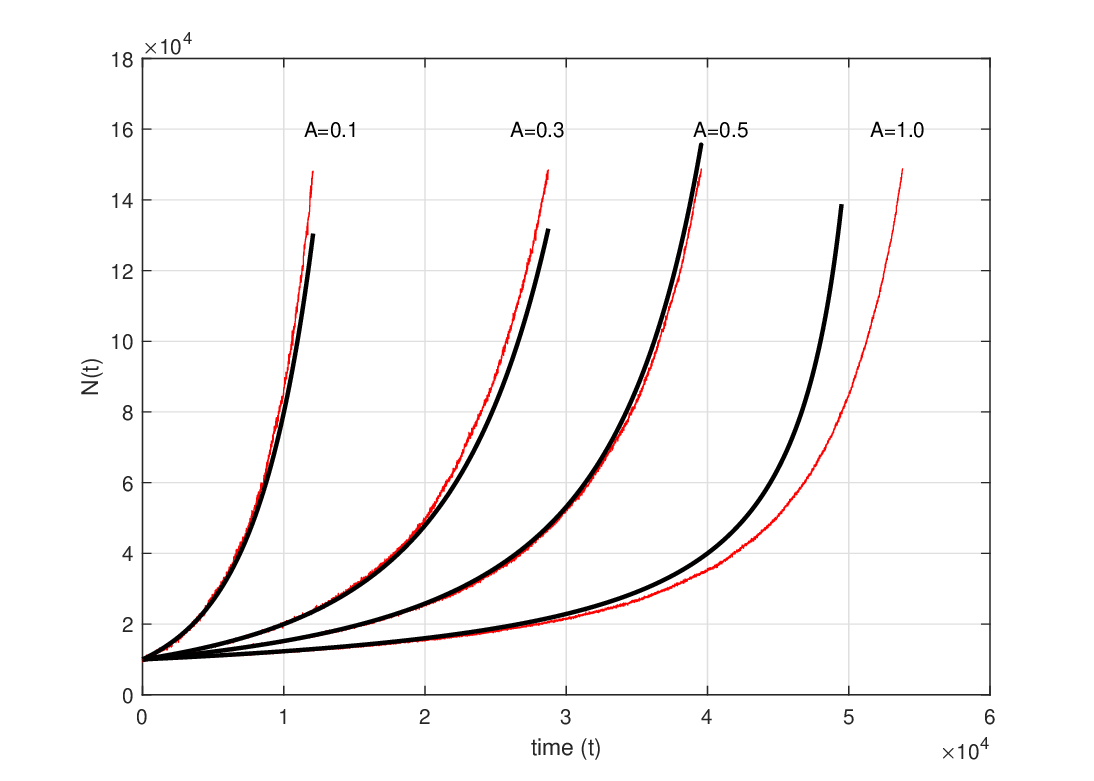}
 \label{fig:5a}
\end{subfigure}
\begin{subfigure}{0.7\textwidth}
\caption{}
\includegraphics[width=\linewidth]{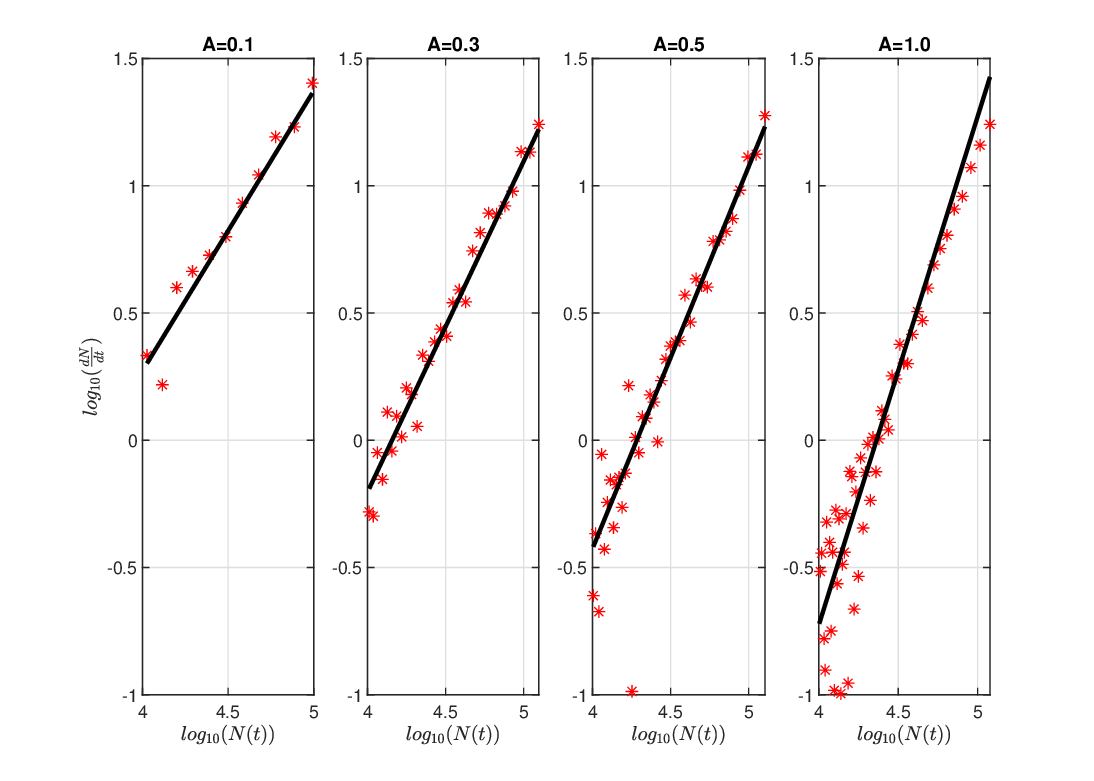}
\label{fig:5b}
\end{subfigure}
\caption{{\bf}
Comparisons of growth patterns of cancer cells population size $N(t)$, for evolution with weakly advantageous mutations, for parameters $f=0.0005$, $p_{f}=0.025$, $N_{C}=10000$ and different values of exponent parameter $A$ in cell death intensity relation (\ref{eq:death_intensity}). Values of $A$ used in computations/simulations are $A=0.1$, $A=0.3$, $A=0.5$ and $A=1.0$. (a): Time plots $N(t)$. Time plots $N(t)$ computed by using analytical relation Eq~(\ref{eq:N_analytical}) are drawn with bold black lines. Time plots of $N(t)$ obtained by using stochastic modelling (Gillespie algorithm) are drawn as red curves. (b): Plots of $log_{10}(\frac{dN}{dt})$ versus $log_{10}(N(t))$. Plots obtained by using relation Eq~(\ref{eq:log_dtN_fit_N}) are drawn as black bold lines. Growth patterns of $N(t)$ obtained on the basis of stochastic simulations (red plots in  Fig~\ref{fig:5a}) are represented by red asterisks. The coordinates of each asterisk are computed as base $10$ logarithms of averaged values of $N(t)$ (horizontal coordinate) and $\frac{dN}{dt}$ (vertical coordinate). Averaging over bins of the size $1~000$ in the time scale is done for the purpose of reducing the large variation of $\frac{dN}{dt}$ in stochastic simulations. 
\label{fig5}
}
\end{figure}

We show the comparison of results of modelling the growth of the population size $N(t)$, by using a deterministic analytical solution Eq~(\ref{eq:N_analytical}) versus stochastic Gillespie algorithm, for parameters $f=0.0005$, $p_{f}=0.025$, $N_{C}=10000$ and different values of exponent parameter $A$ ($A=0.1$, $A=0.3$, $A=0.5$ and $A=1.0$) in cell death intensity relation (\ref{eq:death_intensity}). In Fig~\ref{fig:5a} we present time plots $N(t)$, while in  Fig~\ref{fig:5b} we show plots of $log_{10}(\frac{dN}{dt})$ versus $log_{10}(N(t))$. Black bold curves in  Fig~\ref{fig:5a} and Fig~\ref{fig:5b} are computed by using analytical solution Eq~(\ref{eq:N_analytical}) with approximation Eq~(\ref{eq:sigma_square_fit_const}), where the variance is assumed to equal for all plots, $\sigma_{f}^{2}=25$ (this value was set on the basis of plots shown in  Fig~\ref{fig4} corresponding to $f=0.0005$). Logarithmic plots in Fig~\ref{fig4} B demonstrate that for the range of change of population size $N(t)$ of the order of magnitude $1-2$, the pattern of growth is well approximated by the power law (\ref{eq:log_dtN_fit_N}).

\subsection{Stochastic simulations with Gillespie algorithm \label{subsec:Stochastic-simulations}}

In the subsections above, we have derived deterministic results concerning travelling mutation waves, their shapes (widths) and velocities, and their influence on the cancer population size. These results were verified by using stochastic simulations. The elaborated stochastic simulations algorithm is described here. We use Gillespie stochastic simulation method \cite{Gillespie_1976} to draw random times of events of cells deaths, divisions and mutations, based on intensities Eq~(\ref{eq:death_intensity}), Eq~(\ref{eq:birth_fit}) and probabilities Eq~(\ref{eq:muts_inten_fit}). 
In our implementation of the Gillespie algorithm, the state of the process is $N$ - dimensional vector, where $N$ is the number of cells (population size), each element corresponds to one cell and its entry is determined by the number $l$ of mutations harboured by the cell. Initially, the state vector contains entries equal to $0$ since the initial mutation number is zero for all cells. The initial value of $N$ is set to $N=N_C$.

In each simulation cycle/loop, time instants corresponding to potentially occurring events of deaths and births are randomly drawn from exponential distribution and scaled by intensities Eq~(\ref{eq:death_intensity}), Eq~(\ref{eq:birth_fit}) following from population size and numbers of mutations in each cell. For each cell birth, a possible mutation event is generated randomly as a Bernoulli trial with probability $p_f$. To increase efficiency, we use tau - leap version of the Gillespie algorithm \cite{Marchetti_et_al_book_2017}. The parameter of the simulation algorithm $\tau$ should be much smaller than the unit of time. Based on simulation experiments, we consider the range $0.005 \le \tau \le 0.05$ as a reasonable choice.

Values of randomly generated times of potential deaths and births and results of Bernoulli trials are then used to obtain binary vectors of events (pdt - death event, pdv - division event, and pdm - mutation event), which allow for appropriate updating of the state vector.

The whole simulation process is combined from multiple simulation loops where each loop increases simulation time by $\tau$. Simulation terminates when the condition of simulation time or population size is encountered.

Our implementation of the Gillespie simulation algorithm described above is available at \url{https://pypi.org/project/seEvo1D/}

\subsection{Pipeline of bioinformatic analysis of breast cancer sequencing data\label{subsec:Bioinformatic-pipeline-for}}

We analysed DNA sequencing data on the growth of breast cancer cells populations  
publicly available in NCBI Sequence Read Archive under accession SRP116771 (\url{https://www.ncbi.nlm.nih.gov/sra}). Analyzed dataset concerned 10 synchronous patients samples with both ductal carcinoma in situ (DCIS) and invasive ductal carcinoma (INV) regions sequenced by using topographic single cell sequencing (TSCS) technique \cite{Casasent_et_al_2018}. The dataset consists of 1~959 samples of the single-cell whole genome (WGS) DNA single-end sequencing and 30 samples of whole exome (WXS) DNA paired-end sequencing. Library preparation for sequencing for single cell (SC), WGS data contains whole genome amplification and a random selection, while WXS data includes exome sequencing and hybrid selection. All data was downloaded in FASTQ file format. The number of single cells analyzed for patients varies between 115 to 368 and, when further partitioned according to the type of disease - from 78 to 118, respectively. The WXS data contains one sample for each patient for DCIS, INV, and control (normal cells).
We have started the analysis with bulk data, normal, and tumour samples. We applied quality control of the data by using FaastQC \cite{Andrews_2010} and MultiQC \cite{Ewels_et_al_2016} tools. After the quality check, we performed data cleaning using Trimmomatic \cite{Bolger_et_al_2014}. Then we carried out one-by-one alignment, sorting, compressing, indexing, and marking duplication for reads of the data using Bowtie2 \cite{Langmead_Salzberg_2012} and Samtools \cite{Li_et_al_2009} environments. As a result, we received BAM files with corresponding BAM. BAI files for each sample. We applied (using BAM files for normal samples)  GATK pipeline to generate a panel of normal (PoN), which helps to remove common artifactual and germline variant sites. Creating PoN is a crucial step in GATK best practice workflow for high-confidence somatic mutation calling. After that, we executed Mutect2 to detect and filter all confident somatic mutations for tumour WXS samples, saved as a \textit{vcf} file.  As the last step of the pipeline, we applied Variant Effect Predictor \cite{McLaren_et_al_2016} to predict the potential impacts of the detected somatic mutations. All information about a prediction for each mutation is added to the \textit{vcf} file as a consequence (CSQ) entry in the INFO column. 
For all SC data, we performed the analysis pipeline similar to that of bulk data, obtaining BAM files after quality control, trimming, sorting, compressing, and indexing. One difference between SC and bulk sequencing pipelines was omitting the step of removing duplicates in the SC pipeline. After that, we performed calling of somatic mutations using Mutect2 and PoN previously used for bulk data. Considering the use of different sequencing techniques, we expect PoN, in this case, to help catch only germline variant sites while omitting the artifactual variant sites. 

Results of our analyses, 20 \textit{vcf} files for WXS sequencing of bulk data and 1~959 \textit{vcf} files of WGS sequencing of SC data are available at our repository 
\url{https://doi.org/10.5281/zenodo.7613199}.

\section{Results}

In this section we are confronting our deterministic and stochastic modelling results against some published experimental data. At the beginning we perform a search for driver mutations available in the OncoVar database recently published by \cite{wang2021oncovar} in the whole exome sequencing data from the TCGA atlas  \cite{Wilks_et_al_2014}. Then we use two different experimental studies on the growth of cancer cells population and the related datasets. The first study and experimental dataset related to observed growth of the tumour was published in the paper \cite{Casasent_et_al_2018} on clonal invasion in breast tumours as the scenario of transformation from an early stage (ductal carcinoma in situ, DCIS) to invasive form (invasive ductal carcinoma, INV) in breast cancer, studied by using DNA sequencing techniques. The second study \cite{PerezGarcia_et_al_2020} was devoted to experimental measurements of rates of growth of tumours (cancer cell populations) estimated by measuring intensities of glucose intakes by using PET (positron emission tomography)  technique.

\subsection{Searching for somatic driver mutations in the \textit{vcf} files in the TCGA atlas}
The results shown in this subsection are based upon data generated by the TCGA Research Network: 
\url{http://cancergenome.nih.gov/}, \cite{Wilks_et_al_2014}. We have downloaded \textit{vcf} (variant calling format) files holding lists of all somatic mutations, corresponding to whole exome sequenced samples in the TCGA atlas. The lists of somatic mutations in the downloaded files, obtained by using Mutect2 somatic mutations caller \cite{VanderAuvera_et_al_2013}, were compared to the database of driver mutations OncoVar \cite{wang2021oncovar}. The recently published OncoVar database contains over $20$ thousand driver mutations (variants) detected by the authors in the TCGA atlas data files, on the basis of carefully designed, multistage bioinformatic pipeline. Our comparison is done by using genomic coordinates of somatic mutations (described by GRCh38 reference genome). 
Table~\ref{table_tcga_drivers} shows result of searching for driver mutations detected in whole exome sequenced data in the TCGA atlas in the OncoVar database. For each of the cancer types in the TCGA atlas we list the number of all exome sequenced samples versus the number of samples for which no driver mutation were found in the OncoVar database.

\begin{table}[!ht]
\begin{adjustwidth}{-2.25in}{0in} 
\centering
\caption{
{\bf Results of searching for driver mutations from the OncoVar database in the \textit{vcf} files containing somatic mutations of exome sequenced samples of patients from the TCGA atlas. Numbers of all whole exome sequenced samples from the TCGA atlas versus numbers of samples with no somatic mutations found in the OncoVar database (no driver mutations detected) are listed separately for each of the cancer types.}}
\begin{tabular}{|l|c|c|}
\hline
  \textbf{Cancer type } &                       \textbf{Number of }    &         \textbf{Number of samples with } \\
   &         \textbf{all samples}                  &         \textbf{no driver mutations detected} 
  \\ \thickhline \rowcolor[HTML]{EFEFEF}
Acute Myeloid Leukemia        &                      182        &                        138 \\ \hline
Adrenocortical Carcinoma        &                    92               &                  87 \\ \hline \rowcolor[HTML]{EFEFEF}
Bladder Urothelial Carcinoma                &        426             &                   122 \\ \hline
Brain Lower Grade Glioma            &                530                   &             58 \\ \hline \rowcolor[HTML]{EFEFEF}
Breast invasive carcinoma                 &          1~080             &                  540 \\ \hline
Cervical squamous cell carcinoma  and endocervical adenocarcinoma            &      307                 &               171 \\ \hline \rowcolor[HTML]{EFEFEF}
Colon Adenocarcinoma             &                   179             &                   172 \\ \hline
Diffuse Large B-cell Lymphoma          &             46              &                   38 \\ \hline \rowcolor[HTML]{EFEFEF}
Esophageal carcinoma                 &               185             &                   103 \\ \hline
Glioblastoma multiforme            &                 498                   &             243 \\ \hline \rowcolor[HTML]{EFEFEF}
Head and Neck squamous cell carcinoma         &      512              &                 191 \\ \hline
Kidney renal clear cell carcinoma            &       376               &                 316 \\ \hline \rowcolor[HTML]{EFEFEF}
Kidney renal papillary cell carcinoma       &       293                 &               224 \\ \hline
Liver hepatocellular carcinoma             &         378              &                  233 \\ \hline \rowcolor[HTML]{EFEFEF}
Lung adenocarcinoma                 &                670            &                  153 \\ \hline
Lung squamous cell carcinoma             &           561             &                  208 \\ \hline \rowcolor[HTML]{EFEFEF}
Ovarian serous cystadenocarcinoma            &       610 &                                371 \\ \hline
Pancreatic adenocarcinoma                   &        184             &                  38 \\ \hline \rowcolor[HTML]{EFEFEF}
Pheochromocytoma and Paraganglioma         &         183            &                    149 \\ \hline
Prostate adenocarcinoma                    &         503           &                     383 \\ \hline \rowcolor[HTML]{EFEFEF}
Rectum Adenocarcinoma                      &         164          &                      50 \\ \hline
Sarcoma                                   &          258    &                           231 \\ \hline \rowcolor[HTML]{EFEFEF}
Skin Cutaneous Melanoma                   &          472   &                             22 \\ \hline
Stomach adenocarcinoma                     &         441      &                          166 \\ \hline \rowcolor[HTML]{EFEFEF}
Testicular Germ Cell Tumors               &          156        &                       127 \\ \hline
Thymoma                                &             123           &                     118 \\ \hline \rowcolor[HTML]{EFEFEF}
Thyroid carcinoma                     &              504          &                      141 \\ \hline
Uterine Carcinosarcoma                 &             57            &                    19 \\ \hline \rowcolor[HTML]{EFEFEF}
Uterine Corpus Endometrial Carcinoma    &            561               &                 34 \\ \hline
Uveal Melanoma                          &            79             &                    4 \\ \hline \rowcolor[HTML]{EFEFEF}
\end{tabular}

\label{table_tcga_drivers}
\end{adjustwidth}
\end{table}

As seen from the data in Table~\ref{table_tcga_drivers}, substantial fraction of samples for each of the cancer types lack known driver mutations, despite malignancy of the tumour. This supports the hypothesis that some stages of the tumour growth may be driven by the accumulated effect of many passenger somatic mutations of low frequencies and weak advantageous impact.

\subsection{Data on transition from early stage to the invasive form of breast cancer}

Majority of available cancer DNA sequencing data concern bulk measurements at single time instants. The unique character of the study \cite{Casasent_et_al_2018} was following the progression of breast cancer from the early-stage, ductal carcinoma in situ (DCIS) to the invasive form, invasive carcinoma (INV), combined with single cell DNA sequencing. The authors developed an original experimental technique, topographic single-cell sequencing (TSCS) to measure genomic copy number profiles of single tumour cells. They have applied their experimental technique to $10$ breast cancer patients who experienced DCIS to INV progression. By using TSCS they have identified clonal structures of DCIS and INV tumours defined by copy number variants in the DNA of cancer cells and the direct linkages between clonal structures of DCIS and INV tumours in each of the $10$ patients. They have also performed deep exome bulk sequencing of DCIS and INV cancer samples for all patients.

Among the scientific aims of the paper \cite{Casasent_et_al_2018} was the search for the genomic background behind the progression from DCIS to INV stages of breast cancer.  However, despite exhaustive research, they could not assign cancer progression from DCIS to INV to any exome DNA modification that could be called driver mutation. All discovered mutations occurred at similar allelic frequencies in DCIS and INV samples. They concluded their study with the hypothesis that the genomic factors behind DCIS to INV progression must have occurred prior to the invasion. 

We suppose that the invasion (progression from DCIS to INV) can result from the scenario studied in the present paper, i.e., from the tumour growth with cellular replications stimulated by the occurrence of a large number of weakly advantageous passenger mutations. We have analyzed both bulk exome sequencing and single-cell sequencing data, available in the NCBI Sequence Read Archive, coming from the study  \cite{Casasent_et_al_2018} by using the data processing pipeline described in the subsection above. For all $10$ patients mentioned in the paper, we have identified somatic mutations in bulk sequencing data and in single cells DNA, by using Mutect2 somatic mutations caller \cite{VanderAuvera_et_al_2013}. Below we summarize arguments, which can support the hypothesis that the progression from DCIS to INV stage of breast cancers may be due to the occurrence of a large number of weakly advantageous passenger mutations. 

 In Table~\ref{table_2}, Table~\ref{table_3} we report the number of mutations detected by Mutect2 somatic mutations caller \cite{VanderAuvera_et_al_2013} for bulk sequencing data for all patients in the study. Table~\ref{table_2} shows numbers of all somatic mutations (mutations attributed with PASS status in Mutect2), while Table~\ref{table_3} reports numbers of mutations attributed with PASS status and additionally classified as 'missense variants' by VEP effects predictor \cite{McLaren_et_al_2016}. In Table~\ref{table_2}, Table~\ref{table_3} one can observe that for all somatic mutations $7$ of $10$ patients experience growth of somatic mutation numbers between DCIS and INV stages of cancer, while for the subset of somatic mutations classified as 'missense variants', in $9$ of $10$ patients, there is an increase of mutation numbers between DCIS and INV stages of cancer. The fact that in the majority of patients growth of numbers of mutations is observed would suggest a weak positive selective effect of at least some somatic mutations. Somatic mutations corresponding to 'missense variants' are more likely to cause selective effects.

\begin{table}[!ht]
\centering
\caption{
{\bf Numbers of mutations detected by Mutect2 somatic mutations caller (attributed with PASS status). Numbers of all mutations with PASS status.}}
\begin{tabular}{|c|r|r|r|}
\hline
    & \textbf{DCIS} & \textbf{INV} & \textbf{Intersection of DCIS and INV} \\ \thickhline
   \rowcolor[HTML]{EFEFEF}
P1  & 20 862         & 17 742        & 28                                    \\ \hline
P2  & 10 120         & 15 710        & 65                                    \\ \hline \rowcolor[HTML]{EFEFEF}
P3  & 5 619          & 20 775        & 46                                    \\ \hline
P4  & 986           & 846          & 105                                   \\ \hline \rowcolor[HTML]{EFEFEF}
P5  & 2 335          & 45 216        & 134                                   \\ \hline
P6  & 14 198         & 13 227        & 181                                   \\ \hline \rowcolor[HTML]{EFEFEF}
P7  & 3 148          & 36 072        & 279                                   \\ \hline
P8  & 5 959          & 19 729        & 24                                    \\ \hline \rowcolor[HTML]{EFEFEF}
P9  & 7 046          & 11 660        & 217                                   \\ \hline
P10 & 18 088         & 26 861        & 204                                   \\ \hline \rowcolor[HTML]{EFEFEF}
\end{tabular}

\label{table_2}
\end{table}

\begin{table}[!ht]
\centering
\caption{
{\bf Numbers of mutations detected by Mutect2 somatic mutations caller (attributed with PASS status). Numbers of mutations with PASS status and with the consequence 'missense variant' predicted by VEP predictor. }}
\begin{tabular}{|c|r|r|r|}
\hline
    & \textbf{DCIS} & \textbf{INV} & \textbf{Intersection of DCIS and INV} \\ \thickhline \rowcolor[HTML]{EFEFEF}
P1  & 2 437          & 3 833         & 2                                     \\ \hline
P2  & 1 850          & 3 475         & 13                                    \\ \hline \rowcolor[HTML]{EFEFEF}
P3  & 1 084          & 3 103         & 8                                     \\ \hline
P4  & 170           & 146          & 24                                    \\ \hline \rowcolor[HTML]{EFEFEF}
P5  & 316           & 14 832        & 33                                    \\ \hline
P6  & 3 472          & 3 829         & 51                                    \\ \hline \rowcolor[HTML]{EFEFEF}
P7  & 600           & 11 357        & 65                                    \\ \hline
P8  & 717           & 2 686         & 3                                     \\ \hline \rowcolor[HTML]{EFEFEF}
P9  & 1 205          & 3 006         & 30                                    \\ \hline
P10 & 2 854          & 7 849         & 42                                    \\ \hline \rowcolor[HTML]{EFEFEF}
\end{tabular}

\label{table_3}
\end{table}

Single-cell DNA sequencing data posted by \cite{Casasent_et_al_2018}, give a chance to observe mutation waves by estimating sizes of classes $n_{l}$ (Table~\ref{table_1}). We have therefore searched (by chromosomal positions) for all mutations discovered by Mutect2 in bulk sequencing files, in single-cell DNA sequences, for all patients. Results are reported in Table~\ref{table_4}, for all somatic mutations and in Table~\ref{table_5} for somatic mutations corresponding to 'missense variants'. One can see that both in Table~\ref{table_4} and in Table~\ref{table_5}, for the reported classes of cells $n_{l}$, values of $l$ are very low, single cells in the study harbour very small numbers of somatic mutations detected in bulk sequencing data. This is because DNA coverage in single cells is very low. Nevertheless, for all somatic mutations (Table~\ref{table_4}), in $7$ out of $10$ patients, and for somatic mutations corresponding to 'missense variants' (Table~\ref{table_5}), in $6$ out of $10$ patients, we observe progressions of mutation waves towards increasing values of $\chi_{f}.$ Moreover, in $5$ cases in both tables application of the chi-square test of independence for comparing distributions of cell classes in DCIS and INV cancers for patients, allows for proving the statistical significance of the increase of the value of $\chi_{f}.$

\begin{table}[!ht]
\begin{adjustwidth}{-2.25in}{0in} 
\centering
\caption{
{\bf Sizes of mutation classes (cells with equal numbers of mutation counts) corresponding to all somatic mutations detected by Mutect2 somatic mutations caller (attributed with PASS status).   }}
\begin{tabular}{|l|l|lllllllllllllll|l|l|}
\hline
\textbf{Patient} & {\textbf{\begin{tabular}[c]{@{}l@{}}Cancer\\ type\end{tabular}}} & \multicolumn{15}{l|}{\textbf{Size of class n\_l for mutations count l}}                                                                           & \textbf{mean(l)} & \textbf{p\_val} \\ 
 \cline{3-17} 
                 &                                                                                 & \multicolumn{1}{r|}{\textbf{l:}} & \multicolumn{1}{l|}{\textbf{0}} & \multicolumn{1}{l|}{\textbf{1}} & \multicolumn{1}{l|}{\textbf{2}} & \multicolumn{1}{l|}{\textbf{3}} & \multicolumn{1}{l|}{\textbf{4}} & \multicolumn{1}{l|}{\textbf{5}} & \multicolumn{1}{l|}{\textbf{6}} & \multicolumn{1}{l|}{\textbf{7}} & \multicolumn{1}{l|}{\textbf{8}} & \multicolumn{1}{l|}{\textbf{9}} & \multicolumn{1}{l|}{\textbf{10}} & \multicolumn{1}{l|}{\textbf{11}} & \multicolumn{1}{l|}{\textbf{12}} & \textbf{13} &                  &                 \\ \thickhline \rowcolor[HTML]{EFEFEF}
P1               & DCIS                                                                            & \multicolumn{1}{l|}{}            & \multicolumn{1}{l|}{18}         & \multicolumn{1}{l|}{8}          & \multicolumn{1}{l|}{7}          & \multicolumn{1}{l|}{2}          & \multicolumn{1}{l|}{1}          & \multicolumn{1}{l|}{0}          & \multicolumn{1}{l|}{1}          & \multicolumn{1}{l|}{}           & \multicolumn{1}{l|}{}           & \multicolumn{1}{l|}{}           & \multicolumn{1}{l|}{}            & \multicolumn{1}{l|}{}            & \multicolumn{1}{l|}{}            &             & 1.0270            &                 \\ \hline 
P1               & INV                                                                             & \multicolumn{1}{l|}{}            & \multicolumn{1}{l|}{35}         & \multicolumn{1}{l|}{21}         & \multicolumn{1}{l|}{15}         & \multicolumn{1}{l|}{2}          & \multicolumn{1}{l|}{4}          & \multicolumn{1}{l|}{1}          & \multicolumn{1}{l|}{}           & \multicolumn{1}{l|}{}           & \multicolumn{1}{l|}{}           & \multicolumn{1}{l|}{}           & \multicolumn{1}{l|}{}            & \multicolumn{1}{l|}{}            & \multicolumn{1}{l|}{}            &             & 1.0000                & 0.6952          \\ \hline \rowcolor[HTML]{EFEFEF}
P2               & DCIS                                                                            & \multicolumn{1}{l|}{}            & \multicolumn{1}{l|}{49}         & \multicolumn{1}{l|}{27}         & \multicolumn{1}{l|}{7}          & \multicolumn{1}{l|}{4}          & \multicolumn{1}{l|}{1}          & \multicolumn{1}{l|}{}           & \multicolumn{1}{l|}{}           & \multicolumn{1}{l|}{}           & \multicolumn{1}{l|}{}           & \multicolumn{1}{l|}{}           & \multicolumn{1}{l|}{}            & \multicolumn{1}{l|}{}            & \multicolumn{1}{l|}{}            &             & 0.6477           &                 \\ \hline
P2               & INV                                                                             & \multicolumn{1}{l|}{}            & \multicolumn{1}{l|}{48}         & \multicolumn{1}{l|}{30}         & \multicolumn{1}{l|}{12}         & \multicolumn{1}{l|}{4}          & \multicolumn{1}{l|}{}           & \multicolumn{1}{l|}{}           & \multicolumn{1}{l|}{}           & \multicolumn{1}{l|}{}           & \multicolumn{1}{l|}{}           & \multicolumn{1}{l|}{}           & \multicolumn{1}{l|}{}            & \multicolumn{1}{l|}{}            & \multicolumn{1}{l|}{}            &             & 0.7021           & 0.6828          \\ \hline \rowcolor[HTML]{EFEFEF}
P3               & DCIS                                                                            & \multicolumn{1}{l|}{}            & \multicolumn{1}{l|}{43}         & \multicolumn{1}{l|}{8}          & \multicolumn{1}{l|}{3}          & \multicolumn{1}{l|}{}           & \multicolumn{1}{l|}{}           & \multicolumn{1}{l|}{}           & \multicolumn{1}{l|}{}           & \multicolumn{1}{l|}{}           & \multicolumn{1}{l|}{}           & \multicolumn{1}{l|}{}           & \multicolumn{1}{l|}{}            & \multicolumn{1}{l|}{}            & \multicolumn{1}{l|}{}            &             & 0.2593           &                 \\ \hline
P3               & INV                                                                             & \multicolumn{1}{l|}{}            & \multicolumn{1}{l|}{45}         & \multicolumn{1}{l|}{32}         & \multicolumn{1}{l|}{14}         & \multicolumn{1}{l|}{6}          & \multicolumn{1}{l|}{1}          & \multicolumn{1}{l|}{0}          & \multicolumn{1}{l|}{1}          & \multicolumn{1}{l|}{}           & \multicolumn{1}{l|}{}           & \multicolumn{1}{l|}{}           & \multicolumn{1}{l|}{}            & \multicolumn{1}{l|}{}            & \multicolumn{1}{l|}{}            &             & 0.8889           & 0.0031          \\ \hline \rowcolor[HTML]{EFEFEF}
P4               & DCIS                                                                            & \multicolumn{1}{l|}{}            & \multicolumn{1}{l|}{39}         & \multicolumn{1}{l|}{6}          & \multicolumn{1}{l|}{}           & \multicolumn{1}{l|}{}           & \multicolumn{1}{l|}{}           & \multicolumn{1}{l|}{}           & \multicolumn{1}{l|}{}           & \multicolumn{1}{l|}{}           & \multicolumn{1}{l|}{}           & \multicolumn{1}{l|}{}           & \multicolumn{1}{l|}{}            & \multicolumn{1}{l|}{}            & \multicolumn{1}{l|}{}            &             & 0.1333           &                 \\ \hline
P4               & INV                                                                             & \multicolumn{1}{l|}{}            & \multicolumn{1}{l|}{62}         & \multicolumn{1}{l|}{17}         & \multicolumn{1}{l|}{4}          & \multicolumn{1}{l|}{1}          & \multicolumn{1}{l|}{1}          & \multicolumn{1}{l|}{}           & \multicolumn{1}{l|}{}           & \multicolumn{1}{l|}{}           & \multicolumn{1}{l|}{}           & \multicolumn{1}{l|}{}           & \multicolumn{1}{l|}{}            & \multicolumn{1}{l|}{}            & \multicolumn{1}{l|}{}            &             & 0.3765           & 0.3275          \\ \hline \rowcolor[HTML]{EFEFEF}
P5               & DCIS                                                                            & \multicolumn{1}{l|}{}            & \multicolumn{1}{l|}{51}         & \multicolumn{1}{l|}{19}         & \multicolumn{1}{l|}{8}          & \multicolumn{1}{l|}{0}          & \multicolumn{1}{l|}{1}          & \multicolumn{1}{l|}{}           & \multicolumn{1}{l|}{}           & \multicolumn{1}{l|}{}           & \multicolumn{1}{l|}{}           & \multicolumn{1}{l|}{}           & \multicolumn{1}{l|}{}            & \multicolumn{1}{l|}{}            & \multicolumn{1}{l|}{}            &             & 0.4937           &                 \\ \hline
P5               & INV                                                                             & \multicolumn{1}{l|}{}            & \multicolumn{1}{l|}{25}         & \multicolumn{1}{l|}{27}         & \multicolumn{1}{l|}{18}         & \multicolumn{1}{l|}{17}         & \multicolumn{1}{l|}{5}          & \multicolumn{1}{l|}{3}          & \multicolumn{1}{l|}{2}          & \multicolumn{1}{l|}{1}          & \multicolumn{1}{l|}{}           & \multicolumn{1}{l|}{}           & \multicolumn{1}{l|}{}            & \multicolumn{1}{l|}{}            & \multicolumn{1}{l|}{}            &             & 1.7143           & 2.7767E-06      \\ \hline \rowcolor[HTML]{EFEFEF}
P6               & DCIS                                                                            & \multicolumn{1}{l|}{}            & \multicolumn{1}{l|}{31}         & \multicolumn{1}{l|}{28}         & \multicolumn{1}{l|}{23}         & \multicolumn{1}{l|}{20}         & \multicolumn{1}{l|}{8}          & \multicolumn{1}{l|}{6}          & \multicolumn{1}{l|}{3}          & \multicolumn{1}{l|}{}           & \multicolumn{1}{l|}{}           & \multicolumn{1}{l|}{}           & \multicolumn{1}{l|}{}            & \multicolumn{1}{l|}{}            & \multicolumn{1}{l|}{}            &             & 1.7983           &                 \\ \hline
P6               & INV                                                                             & \multicolumn{1}{l|}{}            & \multicolumn{1}{l|}{99}         & \multicolumn{1}{l|}{49}         & \multicolumn{1}{l|}{26}         & \multicolumn{1}{l|}{5}          & \multicolumn{1}{l|}{6}          & \multicolumn{1}{l|}{3}          & \multicolumn{1}{l|}{1}          & \multicolumn{1}{l|}{}           & \multicolumn{1}{l|}{}           & \multicolumn{1}{l|}{}           & \multicolumn{1}{l|}{}            & \multicolumn{1}{l|}{}            & \multicolumn{1}{l|}{}            &             & 0.8519           & 7.6098E-07      \\ \hline \rowcolor[HTML]{EFEFEF}
P7               & DCIS                                                                            & \multicolumn{1}{l|}{}            & \multicolumn{1}{l|}{15}         & \multicolumn{1}{l|}{21}         & \multicolumn{1}{l|}{12}         & \multicolumn{1}{l|}{7}          & \multicolumn{1}{l|}{5}          & \multicolumn{1}{l|}{0}          & \multicolumn{1}{l|}{2}          & \multicolumn{1}{l|}{}           & \multicolumn{1}{l|}{}           & \multicolumn{1}{l|}{}           & \multicolumn{1}{l|}{}            & \multicolumn{1}{l|}{}            & \multicolumn{1}{l|}{}            &             & 1.5806           &                 \\ \hline
P7               & INV                                                                             & \multicolumn{1}{l|}{}            & \multicolumn{1}{l|}{17}         & \multicolumn{1}{l|}{17}         & \multicolumn{1}{l|}{25}         & \multicolumn{1}{l|}{17}         & \multicolumn{1}{l|}{12}         & \multicolumn{1}{l|}{5}          & \multicolumn{1}{l|}{3}          & \multicolumn{1}{l|}{2}          & \multicolumn{1}{l|}{1}          & \multicolumn{1}{l|}{1}          & \multicolumn{1}{l|}{0}           & \multicolumn{1}{l|}{0}           & \multicolumn{1}{l|}{1}           &             & 2.4950            & 0.1812          \\ \hline \rowcolor[HTML]{EFEFEF}
P8               & DCIS                                                                            & \multicolumn{1}{l|}{}            & \multicolumn{1}{l|}{119}        & \multicolumn{1}{l|}{26}         & \multicolumn{1}{l|}{4}          & \multicolumn{1}{l|}{}           & \multicolumn{1}{l|}{}           & \multicolumn{1}{l|}{}           & \multicolumn{1}{l|}{}           & \multicolumn{1}{l|}{}           & \multicolumn{1}{l|}{}           & \multicolumn{1}{l|}{}           & \multicolumn{1}{l|}{}            & \multicolumn{1}{l|}{}            & \multicolumn{1}{l|}{}            &             & 0.2282           &                 \\ \hline
P8               & INV                                                                             & \multicolumn{1}{l|}{}            & \multicolumn{1}{l|}{92}         & \multicolumn{1}{l|}{84}         & \multicolumn{1}{l|}{28}         & \multicolumn{1}{l|}{12}         & \multicolumn{1}{l|}{3}          & \multicolumn{1}{l|}{}           & \multicolumn{1}{l|}{}           & \multicolumn{1}{l|}{}           & \multicolumn{1}{l|}{}           & \multicolumn{1}{l|}{}           & \multicolumn{1}{l|}{}            & \multicolumn{1}{l|}{}            & \multicolumn{1}{l|}{}            &             & 0.8584           & 2.2751E-11      \\ \hline \rowcolor[HTML]{EFEFEF}
P9               & DCIS                                                                            & \multicolumn{1}{l|}{}            & \multicolumn{1}{l|}{27}         & \multicolumn{1}{l|}{18}         & \multicolumn{1}{l|}{9}          & \multicolumn{1}{l|}{5}          & \multicolumn{1}{l|}{0}          & \multicolumn{1}{l|}{2}          & \multicolumn{1}{l|}{}           & \multicolumn{1}{l|}{}           & \multicolumn{1}{l|}{}           & \multicolumn{1}{l|}{}           & \multicolumn{1}{l|}{}            & \multicolumn{1}{l|}{}            & \multicolumn{1}{l|}{}            &             & 1.0000               &                 \\ \hline
P9               & INV                                                                             & \multicolumn{1}{l|}{}            & \multicolumn{1}{l|}{44}         & \multicolumn{1}{l|}{23}         & \multicolumn{1}{l|}{9}          & \multicolumn{1}{l|}{2}          & \multicolumn{1}{l|}{1}          & \multicolumn{1}{l|}{}           & \multicolumn{1}{l|}{}           & \multicolumn{1}{l|}{}           & \multicolumn{1}{l|}{}           & \multicolumn{1}{l|}{}           & \multicolumn{1}{l|}{}            & \multicolumn{1}{l|}{}            & \multicolumn{1}{l|}{}            &             & 0.6456           & 0.2388          \\ \hline \rowcolor[HTML]{EFEFEF}
P10              & DCIS                                                                            & \multicolumn{1}{l|}{}            & \multicolumn{1}{l|}{20}         & \multicolumn{1}{l|}{14}         & \multicolumn{1}{l|}{18}         & \multicolumn{1}{l|}{15}         & \multicolumn{1}{l|}{8}          & \multicolumn{1}{l|}{6}          & \multicolumn{1}{l|}{3}          & \multicolumn{1}{l|}{1}          & \multicolumn{1}{l|}{2}          & \multicolumn{1}{l|}{}           & \multicolumn{1}{l|}{}            & \multicolumn{1}{l|}{}            & \multicolumn{1}{l|}{}            &             & 2.2759           &                 \\ \hline
P10              & INV                                                                             & \multicolumn{1}{l|}{}            & \multicolumn{1}{l|}{39}         & \multicolumn{1}{l|}{21}         & \multicolumn{1}{l|}{17}         & \multicolumn{1}{l|}{23}         & \multicolumn{1}{l|}{16}         & \multicolumn{1}{l|}{6}          & \multicolumn{1}{l|}{4}          & \multicolumn{1}{l|}{3}          & \multicolumn{1}{l|}{2}          & \multicolumn{1}{l|}{1}          & \multicolumn{1}{l|}{2}           & \multicolumn{1}{l|}{0}           & \multicolumn{1}{l|}{0}           & 1           & 2.3778           & 0.7999          \\ \hline
\end{tabular}
\begin{flushleft} 
Column mean(l) reports mean values of mutation numbers detected in single cells. Column p\_val contains p values of the chi-square independence test comparing distributions of sizes of mutations classes between DCIS and INV cells for each patient.
\end{flushleft}
\label{table_4}
\end{adjustwidth}
\end{table}

\begin{table}[!ht]
\centering
\caption{
{\bf Sizes of mutation classes (cells with equal numbers of mutation counts) corresponding to the subset of somatic mutations detected by Mutect2
with the consequence 'missense variant' predicted by VEP predictor.}}
\begin{tabular}{|l|l|lllllll|l|l|}
\hline
\textbf{Patient} & \textbf{\begin{tabular}[c]{@{}l@{}}Cancer\\ type\end{tabular}} & \multicolumn{7}{l|}{\textbf{Size of class   n\_l for mutations count l}}                                                                                                                                                & \textbf{mean(l)} & \textbf{p\_val} \\ \cline{3-9}
                                  &                                                                                 & \multicolumn{1}{r|}{\textbf{l:}} & \multicolumn{1}{l|}{\textbf{0}} & \multicolumn{1}{l|}{\textbf{1}} & \multicolumn{1}{l|}{\textbf{2}} & \multicolumn{1}{l|}{\textbf{3}} & \multicolumn{1}{l|}{\textbf{4}} & \textbf{5} &                                   &                                  \\ \hline \rowcolor[HTML]{EFEFEF}
P1                                & DCIS                                                                            & \multicolumn{1}{l|}{}            & \multicolumn{1}{l|}{30}         & \multicolumn{1}{l|}{7}          & \multicolumn{1}{l|}{}           & \multicolumn{1}{l|}{}           & \multicolumn{1}{l|}{}           &            & 0.1892                            &                                  \\ \hline
P1                                & INV                                                                             & \multicolumn{1}{l|}{}            & \multicolumn{1}{l|}{68}         & \multicolumn{1}{l|}{8}          & \multicolumn{1}{l|}{2}          & \multicolumn{1}{l|}{}           & \multicolumn{1}{l|}{}           &            & 0.1538                            & 0.2862                           \\ \hline \rowcolor[HTML]{EFEFEF}
P2                                & DCIS                                                                            & \multicolumn{1}{l|}{}            & \multicolumn{1}{l|}{75}         & \multicolumn{1}{l|}{13}         & \multicolumn{1}{l|}{}           & \multicolumn{1}{l|}{}           & \multicolumn{1}{l|}{}           &            & 0.1477                            &                                  \\ \hline
P2                                & INV                                                                             & \multicolumn{1}{l|}{}            & \multicolumn{1}{l|}{86}         & \multicolumn{1}{l|}{7}          & \multicolumn{1}{l|}{1}          & \multicolumn{1}{l|}{}           & \multicolumn{1}{l|}{}           &            & 0.0957                            & 0.1866                           \\ \hline \rowcolor[HTML]{EFEFEF}
P3                                & DCIS                                                                            & \multicolumn{1}{l|}{}            & \multicolumn{1}{l|}{52}         & \multicolumn{1}{l|}{2}          & \multicolumn{1}{l|}{}           & \multicolumn{1}{l|}{}           & \multicolumn{1}{l|}{}           &            & 0.0370                             &                                  \\ \hline
P3                                & INV                                                                             & \multicolumn{1}{l|}{}            & \multicolumn{1}{l|}{90}         & \multicolumn{1}{l|}{9}          & \multicolumn{1}{l|}{}           & \multicolumn{1}{l|}{}           & \multicolumn{1}{l|}{}           &            & 0.0909                            & 0.2177                           \\ \hline \rowcolor[HTML]{EFEFEF}
P4                                & DCIS                                                                            & \multicolumn{1}{l|}{}            & \multicolumn{1}{l|}{42}         & \multicolumn{1}{l|}{3}          & \multicolumn{1}{l|}{}           & \multicolumn{1}{l|}{}           & \multicolumn{1}{l|}{}           &            & 0.0667                            &                                  \\ \hline
P4                                & INV                                                                             & \multicolumn{1}{l|}{}            & \multicolumn{1}{l|}{82}         & \multicolumn{1}{l|}{3}          & \multicolumn{1}{l|}{}           & \multicolumn{1}{l|}{}           & \multicolumn{1}{l|}{}           &            & 0.0353                            & 0.4173                           \\ \hline \rowcolor[HTML]{EFEFEF}
P5                                & DCIS                                                                            & \multicolumn{1}{l|}{}            & \multicolumn{1}{l|}{67}         & \multicolumn{1}{l|}{12}         & \multicolumn{1}{l|}{}           & \multicolumn{1}{l|}{}           & \multicolumn{1}{l|}{}           &            & 0.1519                            &                                  \\ \hline
P5                                & INV                                                                             & \multicolumn{1}{l|}{}            & \multicolumn{1}{l|}{72}         & \multicolumn{1}{l|}{17}         & \multicolumn{1}{l|}{5}          & \multicolumn{1}{l|}{3}          & \multicolumn{1}{l|}{0}          & 1          & 0.4184                            & 0.0881                           \\ \hline \rowcolor[HTML]{EFEFEF}
P6                                & DCIS                                                                            & \multicolumn{1}{l|}{}            & \multicolumn{1}{l|}{79}         & \multicolumn{1}{l|}{32}         & \multicolumn{1}{l|}{7}          & \multicolumn{1}{l|}{1}          & \multicolumn{1}{l|}{}           &            & 0.4118                            &                                  \\ \hline
P6                                & INV                                                                             & \multicolumn{1}{l|}{}            & \multicolumn{1}{l|}{150}        & \multicolumn{1}{l|}{35}         & \multicolumn{1}{l|}{4}          & \multicolumn{1}{l|}{}           & \multicolumn{1}{l|}{}           &            & 0.2275                            & 0.0368                           \\ \hline \rowcolor[HTML]{EFEFEF}
P7                                & DCIS                                                                            & \multicolumn{1}{l|}{}            & \multicolumn{1}{l|}{46}         & \multicolumn{1}{l|}{13}         & \multicolumn{1}{l|}{3}          & \multicolumn{1}{l|}{}           & \multicolumn{1}{l|}{}           &            & 0.3065                            &                                  \\ \hline
P7                                & INV                                                                             & \multicolumn{1}{l|}{}            & \multicolumn{1}{l|}{58}         & \multicolumn{1}{l|}{32}         & \multicolumn{1}{l|}{9}          & \multicolumn{1}{l|}{1}          & \multicolumn{1}{l|}{1}          &            & 0.5644                            & 0.2501                           \\ \hline \rowcolor[HTML]{EFEFEF}
P8                                & DCIS                                                                            & \multicolumn{1}{l|}{}            & \multicolumn{1}{l|}{145}        & \multicolumn{1}{l|}{4}          & \multicolumn{1}{l|}{}           & \multicolumn{1}{l|}{}           & \multicolumn{1}{l|}{}           &            & 0.0268                            &                                  \\ \hline
P8                                & INV                                                                             & \multicolumn{1}{l|}{}            & \multicolumn{1}{l|}{197}        & \multicolumn{1}{l|}{21}         & \multicolumn{1}{l|}{1}          & \multicolumn{1}{l|}{}           & \multicolumn{1}{l|}{}           &            & 0.1050                             & 0.0245                           \\ \hline \rowcolor[HTML]{EFEFEF}
P9                                & DCIS                                                                            & \multicolumn{1}{l|}{}            & \multicolumn{1}{l|}{55}         & \multicolumn{1}{l|}{6}          & \multicolumn{1}{l|}{}           & \multicolumn{1}{l|}{}           & \multicolumn{1}{l|}{}           &            & 0.0984                            &                                  \\ \hline
P9                                & INV                                                                             & \multicolumn{1}{l|}{}            & \multicolumn{1}{l|}{68}         & \multicolumn{1}{l|}{11}         & \multicolumn{1}{l|}{}           & \multicolumn{1}{l|}{}           & \multicolumn{1}{l|}{}           &            & 0.1392                            & 0.4628                           \\ \hline \rowcolor[HTML]{EFEFEF}
P10                               & DCIS                                                                            & \multicolumn{1}{l|}{}            & \multicolumn{1}{l|}{61}         & \multicolumn{1}{l|}{20}         & \multicolumn{1}{l|}{4}          & \multicolumn{1}{l|}{2}          & \multicolumn{1}{l|}{}           &            & 0.3908                            &                                  \\ \hline
P10                               & INV                                                                             & \multicolumn{1}{l|}{}            & \multicolumn{1}{l|}{81}         & \multicolumn{1}{l|}{38}         & \multicolumn{1}{l|}{15}         & \multicolumn{1}{l|}{0}          & \multicolumn{1}{l|}{1}          &            & 0.5333                            & 0.1009                           \\ \hline
\end{tabular}
\begin{flushleft} 
Column mean(l) reports mean values of mutation numbers detected in single cells. Column p\_val contains p values of the chi-square independence test comparing distributions of sizes of mutations classes between DCIS and INV cells for each patient. 
\end{flushleft}

\label{table_5}
\end{table}

\subsection{Growth rates of tumours measured by positron emission tomography (PET)}

Growth plots shown in  Fig~\ref{fig5}, prove that growth rates of the
size of the cancer cells population can be approximated by power functions with exponent $1+A$. In Fig~\ref{fig5}, we have shown approximations of cancer cells population growth for $1+A$ ranging from $1.1$ to $2$ and for population size increased by more than one order of magnitude. Growth patterns presented in Fig~\ref{fig:5b}, showing growth intensity versus population size in logarithmic scales, can replicate analogous, experimentally measured plots published in the paper \cite{PerezGarcia_et_al_2020}. In figure 1 a-h from \cite{PerezGarcia_et_al_2020}, values of exponents of power laws range from $1.182$ to $1.386$ and ranges of sizes of tumour cell populations cover $1 - 2$ orders of magnitude. These values can be easily reproduced by assigning suitable values to the exponent parameter $A$ in our modelling, confirmed by the Gillespie simulation algorithm.

\section{Discussion and conclusions}

Mathematical and simulation models of asexual evolution with mutation
waves travelling/propagating in populations of cells/organisms have
already been extensively studied. The majority of approaches concern
constant size population, Wright-Fisher or Moran model of evolution.
Classical result \cite{Haigh_1978}, concerning mutations bringing
deleterious effects, is a Poisson-like quasi - stationary distribution of
sizes of mutation classes (mutation front) in a constant-size population. 
In the deterministic model the position of the mutation front is fixed, while stochastic effects 
cause advance of the mutation wave/front by the mechanism called Muller's ratchet \cite{Muller_1932}) 
Later studies developed many quantitative
aspects concerning estimating the speed of advance of the mutation
front \cite{Gordo_Charlesworth_2000}. Several studies analyze, with
the assumption of constant population size, evolution with advantageous
mutations, \cite{Desai_Fisher_2007}, \cite{Uecker_Hermisson_2011},
\cite{Neher_2013}. In the context of advantageous mutations often
the interest of the researchers is in estimating the probability of fixation
or the time to fixation \cite{Desai_Fisher_2007}, \cite{Uecker_Hermisson_2011}.
Neher \cite{Neher_2013} proposed the mathematical model of the process
of fast adaptation driven by advantageous mutations, in the form of
coalescent with multiple mergers. In \cite{Rouzine_et_al_2003} and
\cite{Desai_Fisher_2007} models of evolution with constant population
size and two types of mutations, mildly deleterious and mildly advantageous
are studied. Evolution is described in terms of waves/fronts of fitness,
which summarize counteracting effects of two possible types of mutations.
Studies \cite{Rouzine_et_al_2003} and \cite{Desai_Fisher_2007}
use different modelling techniques, however, they come to quite consistent
results. Waves and mean population fitness, which appear in
their scenarios follow from the combined effect of deleterious and
advantageous mutations and depend on mutation intensities and values
of selection/fitness coefficients and on the population size. 

In \cite{McFarland_et_al_2013},
\cite{McFarland_et_al_2014} (already mentioned in the introduction)
a model of evolution with rare, strongly advantageous driver mutations
and counteracting mildly deleterious passenger mutations is proposed
for cancer cell populations. The size of the cancer cells population
is not assumed constant but follows from the influence of the environment
with a given capacity parameter and from the effects of occurring mutations.
Mildly (weakly) deleterious passenger mutations lead to a slow, gradual
decrease in the fitness of cancer cells, while driver mutations cause
positive selective sweeps and define the clonal structure of the cancer
cell population. The evolution of the size of the cancer cells population
follows from an imbalance between the effects of occurring passenger and driver
mutations. Depending on which force is stronger, the population either
expands or goes to extinction. The fate of the cancer cells population
also depends on the initial population size, above some critical value
expansion occurs, while the population with an initial size below the
critical value would extinct. In the model in \cite{McFarland_et_al_2013},
\cite{McFarland_et_al_2014} the cancer cells population, if expanding,
follows an explosive growth scenario with a growth rate proportional to
the square of the population size.

Here we have elaborated and studied models of the evolution of cancer cells population  analogous to \cite{McFarland_et_al_2013},
\cite{McFarland_et_al_2014}, where we also assume that the size of the population can change in time
under the pressure of the environment of capacity $N_{C}$ and occurring mutations. The difference between our study and \cite{McFarland_et_al_2013}, \cite{McFarland_et_al_2014} is that we assume that mutations have weakly advantageous effect on fitness of cancer cells.
Evolution is driven by cell deaths with the intensity given by the power function
Eq~(\ref{eq:death_intensity}) with exponent parameter $A$, and cell
births with intensity depending on the number of weakly advantageous
mutations harboured by a cell Eq~(\ref{eq:birth_fit}). The function
describing cell death intensity, Eq~(\ref{eq:death_intensity}) is different
(more general) than that in \cite{McFarland_et_al_2013}, \cite{McFarland_et_al_2014}
where $A=1$ was assumed. Allowing different values of $A$ gives
an additional degree of freedom in fitting the model to observations. 

We have formulated a deterministic model of cancer
cell evolution and verified this model by stochastic simulations.
Stochastic simulations were based on the Gillespie algorithm.
Our deterministic model was defined
as a system of differential equations for balances of numbers
of cells and numbers of mutations that occur in the evolution. 
We have added a cutoff condition,
i.e., a modification of balances in equations based
on the heuristic assumption that cell divisions cannot happen in the cell
classes of sizes $n_{l}<1$. The cutoff condition was already applied in deterministic models of asexual evolution with mutation waves travelling/propagating in populations of cells/organisms, e.g., in modeling evolution of fitness of population of RNA viruses \cite{Tsimring_et_al_1996}, \cite{Kessler_et_al_1997}.  Here we use it for modeling propagation of mutation wave in the population with increasing size. 
Solutions to cutoff modified equations Eq~(\ref{eq:master_fit_FPmod})
show reasonably good agreement with the results of stochastic simulations, as demonstrated
in Fig~\ref{fig3}. Mutation wave is quasi - stationary, it shows very slow increase of variance
in time. We also analyzed the relation between variance of the mutation wave and the population size $N$ as illustrated in Fig~\ref{fig4}. The variance of the mutation wave depends on the value of the parameter $f$ (decreases with the increase of the fitness parameter $f$) and shows very small changes versus
changing exponent parameter $A$. 

On the basis of deterministic equations, we derive models for
the dynamics of the evolution of the cancer cells population size and for
the propagation of the mutation wave. The derivation has the following
scenario. First, by summing differential equations sidewise,
we derive a differential equation governing the change of the population
size Eq~(\ref{eq:pop_size_eq_fit}). The solution to this differential
equation, $N(t)$, has two-time scales. Fast time scale change of
population size is related to the effects of the environmental pressure
with capacity $N_{C}$, while the slow time population size change
stems from the increase of mean birth intensity due to the accumulation
of weakly advantageous mutations. Next step is using Eq~(\ref{eq:master_fit})
and Eq~(\ref{eq:pop_size_eq_fit}) for deriving differential equations
for frequencies of cell types Eq~(\ref{eq:freqs_dyn_fit}). Finally,
on the basis of Eq~(\ref{eq:freqs_dyn_fit}) we obtain a differential equation
describing the dynamics of evolution of a mean number of mutations Eq~(\ref{eq:wave_fit}).
This differential equation describing the speed of mutation waves as functions
of mutation probability $p_{f}$ and variance of mutation wave $\sigma_{f}^{2}$,
is analogous to relations derived in several papers in the literature,
\cite{Rouzine_et_al_2003}, \cite{Neher_2013}, called Fisher's equation
\cite{Neher_2013}, or breeder's equation \cite{Heywood_2005}. The
observation of variance $\sigma_{f}^{2}(N)$ changing slowly as a
function of $N$ (Fig~\ref{fig4}) motivates us to use constant approximation
$\sigma_{f}^{2}=const$, Eq~(\ref{eq:sigma_square_fit_const}). With
this assumption, we study the dynamics of the slow change of the population
size. We use the model of the dynamics of a mean number of mutations
(\ref{eq:wave_fit}) to write a differential equation for the slow time
change of the population size Eq~(\ref{eq:log_dtN_fit_N}), and then
we obtain an analytical solution Eq~(\ref{eq:N_analytical}). Using the
constant approximation $\sigma_{f}^{2}=const$ has limitations. When
changes in the population size $N$ are very large, the resulting
increase of $\sigma_{f}^{2}$ would change the dynamics (pattern of
growth). The approximation $\sigma_{f}^{2}=const$ works well for
changes of $N(t)$ within orders of magnitude $1-2$ as illustrated
in Fig~\ref{fig:5a}, Fig~\ref{fig:5b}. It should be stressed that the key element in using
models of propagation of mutation wave Eq~(\ref{eq:wave_fit}) and growth
of the population size Eq~(\ref{eq:N_analytical}) is choosing/estimating
the value of $\sigma_{f}^{2}$ as the function $\sigma_{f}^{2}(N)$.
For the choice of $\sigma_{f}^{2}$ we used numerical solutions to
cutoff modified equations Eq~(\ref{eq:master_fit_FPmod}). 

We have elaborated the above described modelling tools with the eventual
aim of verifying whether they can show consistency with
some of the observed experimental data on cancer growth. 
First argument for accumulated effect of many weakly advantageous passenger mutations behind cancer cells population evolution is comparison of the content of recently published database OncoVar with \textit{vcf} files corresponding to exome sequenced samples from TCGA atlas. As seen from Table~\ref{table_tcga_drivers} there are numerous cancer samples in the TCGA atlas with no known driver mutation behind neoplastic cellular population growth. For each of the cancer types, numbers of samples with no driver mutations identified are surprisingly high, which supports the possibility of the mechanism of tumour growth pushed by many weakly advantageous passenger mutations. The number of cases without any known driver mutations, reported in Table~\ref{table_tcga_drivers} can be overly estimated in view of the fact that new researches may 
add cancer drivers to the OncoVar database. In the study \cite{Campbell_et_al_2020} devoted to
analyses of large data-sets of cancer sequencing samples, only in 5 percent 
of analysed cases no driver mutations were identified. Even this estimate, however, leaves the possibility that some cancers or some stages of cancer evolution can lack driver mutations.

We have also confronted
the elaborated models with experimental data on the progression of
breast cancer from the DCIS to INV stage \cite{Casasent_et_al_2018},
and on intensities of tumour growths measured by using the PET technique
\cite{PerezGarcia_et_al_2020}. Since the paper \cite{Casasent_et_al_2018}
paper did not directly identify the genomic background of the progression from DCIS to INV breast cancer, we argued that it could be due to evolution driven by a large number of weakly beneficial mutations. 
In bulk sequencing data we saw the
increase of numbers of somatic mutations between DCIS and INV cancers
in majority of patients (Table~\ref{table_2} and Table~\ref{table_3}). 
Sequencing data of single
cells DNA, although with very low coverage, also suggest progression
of mutation waves towards increasing the number of mutations. The
hypothesis of the evolution driven by weakly advantageous mutations
is also consistent with little intersections between somatic mutations
seen in DCIS and INV samples of the same patient (Table~\ref{table_2} and Table~\ref{table_3})
intuitively explained by low allellc frequencies of majority of mutations.

By introducing the model of intensity of the cell death process in the
form of power function Eq~(\ref{eq:death_intensity}) with exponent parameter
$A$ we are able to predict the pattern of cancer cells population
growth with the growth rate proportional to $N^{1+A}$, Eq~(\ref{eq:dtN_fit_N}).
This pattern of growth, when choosing values of $A$ in the
range from $0.182$ to $0.386$, is consistent with experimental results from
\cite{PerezGarcia_et_al_2020}.

Concluding, we have elaborated modelling tools for the growth of the cancer
cell population driven by weakly advantageous mutations. The deterministic
model shows agreement with the results of stochastic simulations and allows
us to estimate the width and velocity of the mutation wave, and to predict
the pattern of growth of the population. The results of the modelling are
consistent with some of the available experimental data.

\nolinenumbers

%
%
%
\bibliography{plos}



\end{document}